\documentclass[12pt,a4paper]{article}
\usepackage{cite,color,graphics,amsmath,epsfig,rotating}
\usepackage{latexsym}
\usepackage{amssymb}
\usepackage{bbm}
\usepackage{graphicx}
\usepackage{subfig}
\usepackage{wrapfig}
\usepackage{psfrag} % formule dans dessin
\usepackage{xcolor}
\usepackage{eurosym}
\usepackage{feynmp}
\begin{document}

\setlength{\unitlength}{1mm}
\renewcommand{\arraystretch}{1.4}

%------------------------------------------------------------------------
% new definitions, abreviations, etc
%------------------------------------------------------------------------

\def\micromegas      {{\tt micrOMEGAs}}
\def\suspect     {{\tt Suspect}}
\def\micro{{\tt micrOMEGAs}}
\def\calchep      {{\tt CalcHEP}}
\def\lanhep      {{\tt LanHEP}}

\def\ma{M_A}
\def\ra{\rightarrow}
\def\snr{\tilde{\nu}_R}
\def\lsp{\tilde{\nu}_R}
\def\mlsp{m_{\tilde{\nu}_R}}
\def\snl{\tilde{\nu}_L}
\def\mneut{m_{\tilde{\chi}^0_1}}
\def\mchi{m_{\tilde{\chi}^0_i}}
\def\mneutt{m_{\tilde{\chi}^0_2}}
\def\mneuth{m_{\tilde{\chi}^0_3}}
\def\mneutf{m_{\tilde{\chi}^0_4}}
\def\mchar{m_{\tilde{\chi}^+_1}}
\def\mchart{m_{\tilde{\chi}^+_2}}
\def\msel{m_{\tilde{e}_L}}
\def\mser{m_{\tilde{e}_R}}
\def\mslo{m_{\tilde{\tau}_1}}
\def\mslt{m_{\tilde{\tau}_2}}
\def\msul{m_{\tilde{u}_L}}
\def\msur{m_{\tilde{u}_R}}
\def\msdl{m_{\tilde{d}_L}}
\def\msdr{m_{\tilde{d}_R}}
\def\msto{m_{\tilde{t}_1}}
\def\mstt{m_{\tilde{t}_2}}
\def\msbo{m_{\tilde{b}_1}}
\def\msbt{m_{\tilde{b}_2}}
\def\sw{s_W}
\def\cw{c_W}
\def\ca{\cos\alpha}
\def\cb{c_\beta}
\def\sa{\sin\alpha}
\def\sb{s_\beta}
\def\tb{\tan\beta}
\def\ssi{\sigma^{SI}_{\lsp N}}
\def\si{\sigma^{SI}}
\def\sip{\sigma^{SI}_{\chi p}}
\def\ssd{\sigma^{SD}_{\chi N}}
\def\sd{\sigma^{SD}}
\def\sdp{\sigma^{SD}_{\chi p}}
\def\sdn{\sigma^{SD}_{\chi n}}
\def\msl{M_{\tilde l}}
\def\msq{M_{\tilde q}}
\def\bsg{B(b\rightarrow s\gamma)}
\def\bsmu{B(B_s\rightarrow\mu^+\mu^-)}
\def\btau{R(B\rightarrow\tau\nu)}
\def\Omg{\Omega h^2}
\def\sip{\sigma^{SI}_{\chi p}}
\def\amu{\delta a_\mu}
\def\neuto{\tilde\chi^0_1}
\def\neuti{\tilde\chi^0_i}
\def\neutt{\tilde\chi^0_2}
\def\neuth{\tilde\chi^0_3}
\def\neutf{\tilde\chi^0_4}
\def\chargi{\tilde\chi^+_i}
\def\charg{\tilde\chi^+_1}
\def\chargt{\tilde\chi^+_2}
\def\gluino{\tilde{g}}
\def\ul{\tilde{u}_L}
\def\ur{\tilde{u}_R}
\def\stau{\tilde{\tau}}
\def\sl{\tilde{l}}
\def\sq{\tilde{q}}
\def\bone{B^1}
\def\sneutrino{\tilde\nu}
\def\msnu{m_{\tilde\nu_R}}
\def\anu{A_{\tilde\nu}}
\def\sn{\sin\theta}
\def\mzp{M_{Z_2}}
\def\azz{\alpha_{Z}}
\def\tesix{\theta_{E_6}}
\def\beq{\begin{equation}}
\def\eeq{\end{equation}}
\def\gcinq{\gamma_5}
\def\wino{\tilde{W}}
\def\bino{\tilde{B}}
\def\binop{\tilde{B'}}
\def\cw{c_W}
\def\sw{s_W}

% Def. fuer groesser-ungefaehr:
\newcommand{\gsim}{\;\raisebox{-0.9ex}           {$\textstyle\stackrel{\textstyle >}{\sim}$}\;}

\newcommand{\ablabels}[3]{
  \begin{picture}(100,0)\setlength{\unitlength}{1mm}
    \put(#1,#3){\bf (a)}
    \put(#2,#3){\bf (b)}
  \end{picture}\\[-8mm]
} 

%=======================================================================
% Title
%=======================================================================

\begin{flushright}
   \vspace*{-18mm}
   Date: \today
\end{flushright}
\vspace*{2mm}

\begin{center}

{\Large\bf The right-handed sneutrino as thermal dark matter in U(1) extensions of the MSSM} \\[8mm]

{\large   G.~B\'elanger$^1$, J.~Da Silva$^1$  and 
A.~Pukhov$^2$}\\[4mm]
{\it 1) LAPTH, Univ. de Savoie, CNRS, B.P.110,  F-74941 Annecy-le-Vieux Cedex, France\\
     2) Skobeltsyn Inst. of Nuclear Physics, Moscow State Univ., Moscow 119992, Russia 
}\\[4mm]

\end{center}

\begin{abstract}

We investigate the parameter space of a supersymmetric model with an extended U(1) gauge symmetry 
in which the   RH sneutrino is a  thermal dark matter candidate. In this scenario, annihilation of RH sneutrinos proceeds mainly through Higgs or Z' exchange.
We find that sneutrinos   in the mass range from 50GeV to more than 1 TeV can be consistent with both the WMAP limit and the direct detection upper limits.  Powerful constraints from  new gauge boson searches at the LHC as well as from $\Delta M_s$  are incorporated. 
Depending on the choice of the U(1) charge,  these scenarios will be further probe by direct dark matter searches as well as by Higgs searches at the LHC.
\end{abstract}

\section{Introduction}

The  minimal supersymmetric standard model (MSSM) contains two neutral weakly interacting particles that could be  dark matter (DM) candidates, the
neutralino and the sneutrino. While the neutralino has been extensively studied and remains one of the favorite DM candidates~\cite{Goldberg:1983nd,Ellis:1983ew,Roszkowski:2004jc,Baer:2009bu,Ellis:2010kf}, the left-handed sneutrino 
faces severe problems. The sneutrino coupling to the Z boson induces a cross section for elastic scattering off
nuclei that can exceed the experimental limit by several orders of magnitude~\cite{Falk:1994es}. In particular 
latest bounds from CDMS~\cite{Aprile:2010um} or Xenon~\cite{Ahmed:2009zw} cannot be satisfied even for a LH sneutrino mass above 1TeV.
Furthermore the sneutrino annihilation rate is usually too rapid to provide enough dark matter~\cite{Falk:1994es}.

The observation of neutrino oscillations indicative of massive neutrinos gives
a natural motivation for adding a  right-handed (RH) neutrino to the SM fields.
Extending the MSSM  with RH neutrinos and their supersymmetric partners
 provides then an alternate DM candidate, the   right-handed sneutrino (RHSN).  
The smallness of the neutrino masses is usually explained by introducing Majorana mass terms
and making use of  the see-saw mechanism. The natural scale for the RH neutrinos is generally around $10^{12}$~GeV
so that RH neutrinos are too heavy to play a direct role in physics below the TeV 
scale and so are their supersymmetric partners. Note however that the inverse see-saw mechanism 
proposes scenarios with RH Majorana neutrinos  at the TeV scale~\cite{Mohapatra:1986aw,Mohapatra:1986bd}. We will not consider these scenarios. 
It is also possible to generate neutrino masses through Dirac mass terms at the expense of
introducing  some large hierarchy among the fermions. In this case 
the supersymmetric partners of the neutrinos are  expected to be, as for other
sfermions, at the SUSY breaking scale, i.e. around or below 1TeV. 
In this framework the RH sneutrino can be the lightest supersymmetric particle (LSP). This is the scenario we will consider here. 

To make a RHSN LSP  a viable dark matter candidate requires special conditions. 
The RH sneutrino being sterile under standard model gauge interactions cannot be 
brought into thermal equilibrium.\footnote{Note however that non thermal mechanisms can make a mostly sterile sneutrino
a good  dark matter candidate~\cite{Asaka:2005cn,Asaka:2006fs,Gopalakrishna:2006kr,Yaguna:2008mi}.} Nevertheless several proposals  
for sneutrino dark matter have emerged including 
mixed sneutrinos  ~\cite{ArkaniHamed:2000bq,Borzumati:2000mc,MarchRussell:2009aq,Kumar:2009sf,Belanger:2010cd},
RHSN in models with Dirac mass terms  that result from the decay of thermal equilibrium MSSM particles~\cite{Asaka:2005cn,Asaka:2006fs},
DM from a RH sneutrino condensate~\cite{McDonald:2007mr}, sneutrinos in inverse see-saw models~\cite{Hall:1997ah, Arina:2008bb,An:2011uq,Bandyopadhyay:2011qm} or RHSN in extensions of the MSSM~\cite{Long:2007ev,Arina:2008yh,Cerdeno:2009dv,Demir:2009kc,Cerdeno:2011qv,Kang:2011wb}.
Extending the gauge group provides another alternative as the RHSN can reach thermal
equilibrium because it couples to new  vector and/or scalar fields  ~\cite{Lee:2007mt,Langacker:2008yv,Bandyopadhyay:2011qm}.
For example an additional U(1)' gauge symmetry  provides new couplings of the sneutrino with  
 the Z' as well as to new scalar fields. In this framework annihilation of pairs of  RH sneutrino  
can be efficient enough to obtain $\Omega h^2\approx 0.1$. The annihilation is specially enhanced  when the particle exchanged in the s-channel is near resonance. Furthermore the elastic scattering cross section of the RHSN is naturally suppressed by several orders of
magnitude as compared to the MSSM sneutrino as on the one hand  the couplings to the EW scale particles (the Z and  the light Higgs) are strongly suppressed and on the other hand the Z' exchange is suppressed because its mass is above the TeV scale. This is the framework  that we will study here.

Models with extended gauge symmetries are well motivated  within the context of superstring models~\cite{Cvetic:1995rj}, grand unified 
theories~\cite{Langacker:1980js,Hewett:1988xc} or  little Higgs models~\cite{ArkaniHamed:2002qy,Han:2003wu}. 
Furthermore in its supersymmetric version, the additional U(1)' symmetry provides an elegant solution to the $\mu$ problem. 
Indeed as in the NMSSM~\cite{Ellwanger:2009dp}
the effective $\mu$ parameter is related to the vev of the singlet $S$ responsible for the breaking of  the U(1)' symmetry~\cite{Cvetic:1997ky,Langacker:1998tc}.
Another interesting feature of supersymmetric models with extended U(1) symmetry is that a $\lambda S H_u H_d$ interaction allows 
for the strong first order phase transition that is needed for electroweak baryogenesis~\cite{Ahriche:2010ny}. 
Finally the light Higgs mass is generally above that of the MSSM light Higgs and thus  above the direct limit from LEP~\cite{Barger:2007im}. Indeed in addition to
MSSM-like contributions from squarks and quarks, the light Higgs mass receives  contributions from the superpotential (as in the NMSSM)~\cite{BasteroGil:2000bw} as well as   from U(1)' D terms.

In the MSSM with U(1) extended gauge symmetries (UMSSM), the dark matter candidate can either be the 
neutralino  or the sneutrino. The neutralino was investigated in~\cite{Kalinowski:2008iq,deCarlos:1997yv,Barger:2007nv}.  We rather concentrate on the RHSN LSP. 
The RHSN dark matter was  considered in  singlet extensions of the MSSM~\cite{Cerdeno:2008ep,Cerdeno:2009dv} in hybrid inflationnary models~\cite{Deppisch:2008bp} as well as 
in different set-ups~\cite{Arina:2007tm,Arina:2008yh} or in models with inverse seesaw~\cite{Arina:2008bb}.
Sneutrino dark matter in different  U(1) extensions of the MSSM was also examined in~\cite{Lee:2007mt} 
and in~\cite{Allahverdi:2009ae}. Their signatures in cosmic rays~\cite{Allahverdi:2009ae,Demir:2009kc} or neutrino telescopes~\cite{Allahverdi:2009se} were also explored. 

In~\cite{Lee:2007mt} sneutrinos annihilation trough Z'  exchange  and $\tilde{B}'$ t-channnel exchange
into neutrinos were considered while the interactions through the Higgs sector were included in 
~\cite{Allahverdi:2009ae,Demir:2009kc}.
We extend those analyses by including all possible annihilation and co-annihilation channels 
and  by performing a complete exploration of the  parameter space allowing for different choices of  the U'(1) symmetry within the context of an $E_6$ unified model.
Furthermore we take into account recent LHC results on the Z', on supersymmetric particles and on the Higgs sector. 
Insisting on having a sneutrino LSP, to complement previous studies that had concentrated on the neutralino case,  will lead to some constraint on the parameter space. 
In particular the parameters $\mu$ and $M_1,M_2$ have to be larger than the sneutrino mass in order to avoid a higgsino or a gaugino LSP.
We will generically consider only cases with colored sparticles above the TeV scale to easily avoid LHC limits on squarks. 

Within the UMSSM with a RHSN dark matter candidate, we found a variety of annihilation channels for the sneutrino, with a predominance of annihilations near a resonance. 
The  main annihilation  processes  can be classified as
\begin{itemize}
\item{} Annihilation near a light Higgs resonance
\item{} Annihilation near a heavy Higgs resonance
\item{} Annihilation near a $Z_2$ resonance 
\item{} Annihilation into W/Z pairs through Higgs exchange
\item{} Coannihilation with the NLSP, here the NLSP can be neutralino, chargino, charged slepton or any other sfermion.
\end{itemize}

Note that annihilation into neutrinos can occur either  through  $Z_2$ exchange  or through $\tilde{B}',\tilde{S}$ exchange when the latter are near the mass of the LSP. 
The neutrino channels are however never the dominant ones. Finally
annihilation of sneutrinos into pairs of new gauge bosons could also contribute,  we will not consider this channel here as
it is relevant only for sneutrino heavier than the Z', that is  above the TeV scale. 
For such a heavy sneutrino to be the LSP means that all soft parameters are also above the TeV scale. 
These configurations for supersymmetric particles have a restricted parameter space. For example,  very large values of $\mu$ often lead to a light Higgs mass below the current limits. Furthermore when the sneutrino is heavier than the Z', the neutralino can become the LSP.
We therefore choose rather to concentrate on the subTeV scale sneutrinos. 
For each scenario we also include an analysis of the direct detection rate as the spin independent 
cross section of sneutrinos on nucleons turns out to pose severe constraints on a whole class of U(1)' models. 

This paper is organized as follows. In section 2 we describe the various sectors of the model as well as the main particle physics constraints. Section 3 presents the relic abundance and section 4 the direct
detection rate. Section 5 contains the results for two sample U(1)' models,  the $U(1)_\psi$ and $U(1)_\eta$ models,  as well as results from a global analysis of the parameter space. Our conclusions are contained in Section 6, while details on the Higgs mass matrices are presented in the Appendix.

\section{The model}

The symmetry group of the model is $SU(3)_C \times SU(2)_L \times U(1)_Y \times U(1)'$.
For definiteness, we assume that this  model is derived from an $E_6$ model, 
in this case the most general U(1)' charges of fermions are described
by only one  parameter, $\theta_{E_6}$,
\begin{equation}Q' = \cos \theta_{E_6} Q_{\chi} + \sin \theta_{E_6} Q_{\psi},\end{equation}
where $-\pi/2 \le \theta_{E_6} \le \pi/2$ and $Q_\chi$, $Q_\psi$ are the generators of two U(1) subgroups of $E_6$~\cite{Langacker:2008yv}. 
The charges of the MSSM left-handed fermions  as well as that of the right-handed neutrino are
defined  in Table~\ref{tab:u1}.  

We also assume that the exotic (s)fermions that belong to the
fundamental representation of $E_6$ and that are needed for anomaly cancellation are above the TeV scale.
We  will neglect them in the following. 
The only exception is the RH neutrino/sneutrino. 
We choose the normalization of the coupling constants such that $g_1'=\sqrt{5/3} g_Y$ where 
$g_Y$ and $g'_1$ are the coupling
constants of $U(1)_Y$ and $U(1)'$ respectively and $g_Y=e/\cw$.  
For the detailed phenomenological analysis we will consider two values of the $\tesix$ angle corresponding to   
$\tesix= \pi/2$, $Q_\psi$ and $\tesix=-.29\pi$  ($\tan\tesix=-\sqrt{5/3}$), $Q_\eta$. The  first choice  corresponds to the U(1)
subgroup in the breaking of $E_6\rightarrow SO(10)\times U(1)_\psi$.
The second choice is used as an example of a scenario in which the  sneutrino dark matter candidate differs 
significantly from the case  of the $U(1)_\psi$ model. 
Note that the RH sneutrino decouples when $\tesix=.42\pi$ 
and that for $\theta_{E_6}=0$  the symmetry cannot be broken by the  singlet field, S, as is the case for other U(1)' models.

\begin{table}[!htb]
\begin{center}
\caption{\label{tab6} Charges of the left-handed fermions  under $U'(1)$.}
\begin{tabular}{|c|c|c|c|c|c|c|c|c|c|}
       \hline
       & Q &  $u^c$ & $d^c$ & L & $e^c$ & $\nu^c$ & $H_u$ & $H_d$ & $S$ \\ \hline 
      $\sqrt{40}Q_{\chi}$ & $-1$ & $-1$ & $3$ & $3$ & $-1$ & $-5$ & $2$ & $-2$ & $0$ \\ \hline 
      $\sqrt{24}Q_{\psi}$ & $1$ & $1$ & $1$ & $1$ & $1$ & $1$ & $-2$ & $-2$ & $4$ \\ \hline 
 $2\sqrt{15}Q_{\eta}$    & $-2$ & $-2$ & $1$ & $1$ & $-2$ & $-5$ & $4$ & $1$ & $-5$ \\ \hline 
\end{tabular}
\label{tab:u1}
\end{center}
\end{table}

In addition to the MSSM superfields, the model contains a new vector superfield, $B'$,  and a new singlet scalar superfield, $S$.
The superpotential is the same as in the MSSM with $\mu=0$ but has an additional term involving the singlet,
\begin{equation}
W=W_{MSSM}|_{\mu=0}+ \lambda S H_u H_d
\end{equation}
The  vev  of S, $\langle S \rangle= v_s/\sqrt{2}$  breaks the U(1)' symmetry and induces a $\mu$ term
\begin{equation}
\mu=\lambda v_s/\sqrt{2}.
\label{eq:mu}
\end{equation}
The physical fields of the model are those of the MSSM with in addition a gauge boson, Z', 
and its associated gaugino, $\tilde{B}'$, 
a scalar field, S, and a neutral  higgsino, $\tilde{S}$. This leads altogether to 6 neutralinos and  to a Higgs sector
with   3 CP-even scalars, 1 CP-odd pseudoscalar and a charged Higgs.  
Each sector of the model that will play a role in dark matter annihilation is described below.

\subsection{Gauge bosons}

The two neutral massive gauge bosons, $Z^0$ and $Z'$ can mix both through mass and kinetic
mixing~\cite{Langacker:2008yv,Kalinowski:2008iq}.  
In the following we will neglect the kinetic mixing as it is not directly relevant to our analysis. 
The electroweak and U(1)' symmetries are broken respectively by the vev's of the doublets, 
$v_u/\sqrt{2}=\langle H_u \rangle,v_d/\sqrt{2}=\langle H_d \rangle$ and singlet.
The mass matrix reads
\begin{equation}
  M^2_{Z} =
  \left( \begin{array}{cc}
   M^2_{Z^0} &   \Delta_Z^2\\
   \Delta_Z^2 & M^2_{Z'}
  \end{array}\right) \,.
\label{eq:mz}
\end{equation}
where 
\begin{eqnarray}
M^2_{Z^0} &= &\frac{1}{4} g_1^2 (v_u^2+v_d^2)\nonumber\\
M^2_{Z'} &= &{g'_1}^2 (Q_1^2 v_d^2+ Q_2^2 v_u^2+{Q'}_S^2 v_s^2)
\end{eqnarray}
and  we have defined $Q_1=Q'_{H_d}$, $Q_2=Q'_{H_u}$, and $g_1=e/s_Wc_W$.  Invariance of the superpotential imposes the condition,  $Q'_S=-(Q_1+Q_2)$.
The mixing term 
\begin{equation}
\Delta_Z^2 = - \frac{1}{2} g_1 g_1' (Q_1 v_d^2 -Q_2 v_u^2)
\end{equation}
Diagonalisation of the mass matrix leads to two eigenstates
\begin{eqnarray}
Z_1&=&\cos\azz Z^0 + \sin\azz Z'\nonumber\\
Z_2&=& - \sin\azz Z^0 + \cos\azz Z'
\end{eqnarray}
where the mixing angle is defined as
\begin{equation}
\sin 2 \azz =\frac{2 \Delta_Z^2}{M^2_{Z_2} - M^2_{Z_1}}
\end{equation}
and  masses of the physical fields 
\begin{equation}
M^2_{Z_1,Z_2}=\frac{1}{2} \left( M^2_{Z^0}+M^2_{Z'} \mp \sqrt{\left(M^2_{Z^0}+M^2_{Z'}\right)^2+4\Delta_Z^4}  \right)
\end{equation}.
The mixing angle is constrained from precise measurements of $Z^0$ properties to be of the order or smaller than $10^{-3}$~\cite{Erler:2009jh},
the new gauge boson $Z_2$ will therefore have approximately the same properties as  the $Z'$. 
As input parameters we choose the physical masses, $M_{Z_1}=91.187{\rm GeV}$, $M_{Z_2}$ and the mixing angle, 
$\azz$. From these together with the coupling constants, we extract both the value of  $\tan\beta=v_u/v_d$ and the 
 value of $v_s$. Note that the W mass is a parameter derived from the $Z^0$ boson mass, $M_W=M_{Z_0} \cos\theta_W$.
When $M_{Z_2}\gg M_{Z_1}$ and $\alpha_Z<<1$, the value of $\tan\beta$ is the solution of the following equation
\begin{equation}
\cos^2\beta(-Q_1+Q_2 \tan^2\beta) \simeq    \frac{g_2 \cos\theta_W}{4 g'_1}\sin\alpha_Z \frac{M_{Z_2}^2}{M_W^2}
\end{equation}
For each U(1)' model, the value of $\tan\beta$ is strongly constrained from the lower bound on the $Z_2$ mass and on its mixing.
For example for $U(1)_\psi$ with $\sin\alpha_Z>0$, the value of $\tan\beta$ has to be below 1, this is because  $\Delta^2_{Z} \sim 1-\tan^2\beta$.  Large values of $\tan\beta$ are found when $\sin\alpha_Z<0$. The large values of $\tan\beta$ are constrained as they this might require too large a mixing angle for a given $Z_2$ mass. For $U(1)_\eta$, $\Delta^2_{Z} \sim 4\tan^2\beta-1$, therefore $\tan\beta< 0.5$  when   $\sin\alpha_Z<0$.
One might think that small values of $\tan\beta$ are problematic for the Higgs mass, however as we will see below additional corrections to the light Higgs mass can bring it above 114 GeV.

\subsection{Higgs sector}

The CP even mass matrix is obtained after minimization of the potential. The Higgs sector of the UMSSM is described in ~\cite{Barger:2006dh} . 
We expand the Higgs field as
\begin{align}
H_d^0 & = \frac{1}{\sqrt{2}} \left( v_d + \phi_d + i \varphi_d \right)\\
H_u^0 & = \frac{1}{\sqrt{2}} \left( v_u + \phi_u + i \varphi_u \right)\\
S     & = \frac{1}{\sqrt{2}} \left( v_s + \sigma + i \xi \right),
\end{align} 
The squared mass in the  $\phi_d,\phi_u,\sigma$ basis  both  at tree level
and including  the dominant radiative corrections from top quarks and stops is listed in Appendix A ~\cite{Barger:2006dh}.
This  squared mass  matrix  is diagonalized by the transformation 
$ M_D=Z_h {\cal M} Z_h^{-1}$
and the mass eigenstates (ordered in mass) are 
$(h_1,h_2,h_3)^T=Z_h^T (\phi_d,\phi_u,\sigma)^T$.

There is only one pseudoscalar Higgs which is obtained after diagonalizing the $3\times 3$ 
pseudoscalar mass matrix listed in Appendix A. This matrix is diagonalized by the unitary transformation $M_D= Z_A {\cal M_A} Z^{-1}_A$ with 
the physical pseudoscalar,  
$A^0= Z_{A1} \varphi_d + Z_{A2} \varphi_u  +Z_{A3} \xi$.
The tree-level mass 
\beq
m_A^2= \frac{\sqrt{2} \lambda A_\lambda}{\sin 2\phi} v
\eeq
and $\tan\phi=v \sin 2\beta/2 v_s$. When $v_s>>v$, the pseudoscalar mass reduces to 
\beq
m_A^2\simeq \frac{\mu A_\lambda}{\sin\beta\cos\beta} \label{eq:ma}
\eeq
The full matrix including one-loop corrections is also listed in Appendix A together with the corrections to the
 charged Higgs mass.

Typically the Higgs spectrum will consist of a standard model like light CP-even Higgs,  
 a heavy mostly doublet  scalar which is almost degenerate with the pseudoscalar and the charged Higgs, 
 and a predominantly singlet scalar.  Note that the mass of the pure singlet 
 $m_S\approx g'_1  Q_s v_s$ and is therefore close to that of $M_{Z_2}$ when $v_s \gg v_u,v_d$, see Eq.~\ref{eq:mz}.
 The hierarchy in the  mass of the heavy doublet and the singlet depends on the parameters of the model.
For large values of $A_\lambda$ and $\lambda$ (and therefore $\mu$) the mass of the heavy Higgs doublet  increases and can exceed that of the singlet. 
Furthermore  the Higgs mixing increases, however  the singlet component of the light state is usually not large enough to significantly relax the bound on the lightest Higgs contrary to what occurs in the NMSSM ~\cite{Ellwanger:2009dp,Ellwanger:1999ji,Ellwanger:1995ru,Franke:1995xn}. 
Although the lightest  scalar Higgs is usually standard model like, it  can be significantly heavier than in the MSSM. 
Indeed the upper bound~\cite{Barger:2006dh} on the Higgs receives two types of additional contributions as compared to the MSSM,
one proportional to $A_\lambda$ that is also found in the NMSSM, and one  from the additional gauge coupling $g'_1$. The upper bound on 
the lightest Higgs mass is thus raised to ${\cal O}(170)$~GeV~\cite{Barger:2006dh}. Such a large Higgs mass is however constrained by the latest LHC results
which for standard model couplings exclude  $m_h>144$~GeV at the 90\%C.L.~\cite{CMS_higgs}.  
The impact of the LHC results on the parameter space of the model will be discussed in more details in~\cite{jonathan_prepa}.

\subsection{Sfermions}
The important new feature in the sfermion sector is that the  U(1)' symmetry induces some 
new D-terms contributions to the sfermion mass, $\Delta_f$,
\begin{equation}
\Delta_f= \frac{1}{2} g_1'^2 Q'_f \left( v^2 (Q_1 \cos^2\beta+ Q_2 \sin^2\beta) + Q'_S v_s^2 \right)
\end{equation} 
where $Q'_f$ are the U(1)' charges of the left fields listed in Table~\ref{tab:u1}.   
The sfermion mass matrix reads
\begin{equation}
  m^2_{\tilde f} =
  \left( \begin{array}{cc}
   {m}^2_{\widetilde{L}} + (I^3_f-Q_f s_W^2) M^2_{Z^0} \cos 2\beta +m_f^2 +\Delta_f &  
   m_f( A_f -\mu \tan\beta^{-2I_f^3})\\
   m_f( A_f -\mu \tan\beta^{-2I_f^3}) &  {m}^2_{\widetilde{f_R}}-Q_f s_W^2 M^2_{Z^0} \cos 2\beta +m_f^2 +\Delta_{f^c}
  \end{array}\right) 
\label{eq:sfermion}
\end{equation}
where $Q_f$ and $I_3^f$ are the charge and isospin of the SM fermions.
The new D-term contribution  can completely dominate the sfermion mass, especially for large values of $v_s$. Those are found  in particular when $\tesix\approx 0$. The D-term contribution can induce negative corrections to the mass, so that light sfermions can be found even when the soft masses are set to 2TeV. For $\tesix>0$, $\Delta_\nu < \Delta_{Q,u_R,e_R}<\Delta_{L,d_R} $ so that a universal soft mass term  for the sfermions at the weak scale 
naturally leads to a
RH sneutrino as the lightest sfermion. Furthermore the NLSP will be the right-handed slepton or the t-squarks if a large mixing decreases the
mass of the lightest $\tilde{t}$. 
On the other hand for $\tesix<0$, $\Delta_\nu >\Delta_{Q,u_R,e_R} > \Delta_{L,d_R}$ so the sneutrino cannot be the LSP with universal soft sfermion masses
at the weak scale. In general  
one expects non-universality in sfermion masses at the weak scale even if universality is imposed when the model is embedded in a GUT scale model. 
In particular the right-handed sleptons, whose RGE's are driven only by U(1) couplings have the smallest soft terms at the weak scale. 
Therefore as long as  $\Delta_\nu$ is not much larger than  for other sfermions, it is still natural that the sneutrino 
be the lightest sfermion.
This is the case for models where $\tesix \approx -\pi/2$ since all correction terms  are then almost equal.
For the sake of reducing the number of free parameters we will in this study assume that the soft sfermion masses are universal 
at the weak scale safe for that of the RH sneutrino. In some cases we will find a sfermion NLSP that contributes to coannihilation processes.
In the $U(1)_\eta$ model the NLSP will  be mostly the left-handed slepton or down-type squarks while in the $U(1)_\psi$ model it will be rather  right-handed slepton or up-type squarks.

\subsection{Neutralinos}

The mass terms for the gaugino  sector reads
\begin{equation}
{\cal L}= -\frac{1}{2} M_1 \tilde{W}\tilde{W} -\frac{1}{2} M_1 \bino\bino - \frac{1}{2} M'_1\binop\binop -M_K \bino\binop+h.c. 
\end{equation} 
The neutralino mass matrix in the basis $(\tilde{B},\tilde{W}^3,\tilde{H}_d,\tilde{H_u},\tilde{S},\tilde{B}')$ thus reads
\begin{equation}
 M_{\tilde \chi^0} =
  \left( \begin{array}{cccccc}
   M_1 	&   0    & -M_{Z^0} \cb \sw  &    M_{Z^0} \sb\sw  &     0     &   M_K\\  
   0 	&   M_2    & M_{Z^0}\cb\cw  & -M_{Z^0}\sb\cw  &     0     &   0\\  
   -M_{Z^0}\cb\sw  	&   M_{Z^0}\cb\cw    & 0 &   -\mu  &     -\mu v\sb/v_s     & Q_1 g'_1 v\cb   \\  
    M_{Z^0}\sb\sw 	&   -M_{Z^0}\sb\cw   & -\mu &   0 &     -\mu v\cb/v_s      &   Q_2 g'_1 v\sb \\  
   0 	&   0     &  -\mu v \sb /v_s    &    -\mu v \cb/v_s    & 0 & Q'_S g'_1 v_s\\  
   M_K 	&   0    & Q_1 g'_1 v\cb  &   Q_2 g'_1 v\sb  &  Q'_S g'_1 v_s  &   M'_1\\  
  \end{array}\right) \,.
\label{eq:sfermion}
\end{equation}
The mixing matrix is defined such that the lightest neutralino is written as
\begin{equation}
\tilde\chi^0_1= {N}_{11} \tilde{B} + {N}_{12} \tilde{W^3}  +{N}_{13} \tilde{H_d} +{N}_{14} \tilde{H_u} +{N}_{15} \tilde{S} +{N}_{16}
\tilde{B'} 
\end{equation}
 Note that $v_s>> v$ so that one expects  a large mixing between the singlino  and $\binop$.
 The chargino sector is identical to that of the MSSM. 

\subsection{Constraints on the model}
\label{sec:constraint}
The main constraints on the model arise from the gauge boson and Higgs sector.  Since colored   supersymmetric particles do not play a direct role in sneutrino annihilation, we will generally assume
that they are above the TeV scale thus evading the LHC constraints~\cite{Collaboration:2011zy,Aad:2011ib}.

The $Z_2$ can be produced directly in $pp$ and $p\bar{p}$ collisions and is searched primarily using leptonic decay modes.  
The limits from the LHC released earlier this year have superseded the Tevatron limits~\cite{Erler:2009jh}. 
We will use the ATLAS exclusion limits published in ~\cite{Collaboration:2011dc} obtained with an integrated luminosity of ${\cal L}=1.01(1.21)fb^{-1}$ in the $e^+e^-(\mu^+\mu^-)$ channels. 
These limits are extracted using different U(1)' models, in the two models we will consider for the phenomenological analysis the bounds are similar,  with  $\mzp >1.49$TeV for $U(1)_\psi$ and $\mzp >1.54$~TeV for $U(1)_\eta$. These bounds were derived assuming the $Z_2$ decays only  into SM particles. 
In our case, the $Z_2$ can also decay into supersymmetric particles, into RH neutrinos and into Higgses, thus reducing  the branching ratio into $e,\mu$.  For example in the $U(1)_\psi$ model the decays into supersymmetric particles can reach up to 20\% especially in cases where the sneutrino is light (hence neutralinos can be light as well) while decays into neutrinos are typically below 10\%. In the $U(1)_\eta$, the decays into SM particles is even more suppressed, it  never exceeds 65\%. This is mainly due to a large branching fraction into right-handed neutrinos ( around 30\%) as well as into other sparticles.    
  The limits on the $Z_2$ mass are therefore weakened. 
To take this effect into account we have computed the  modified leptonic branching ratio for each point in our scans and have  re-derived the corresponding ATLAS limit.

Another important constraint on the gauge boson sector comes from 
precision measurements at the Z-pole and from low energy neutral currents. These  provide stringent constraints on the
mixing angle $\alpha_Z$. Depending on the model parameters the constraint are below a few $10^{-3}$~\cite{Erler:2009jh}. We will typically set the mixing angle to $|\alpha_Z|=0.001$.

The Higgs sector is also severely constrained, a standard model like Higgs is constrained by LEP to be above 114 GeV~\cite{Nakamura:2010zzi} and by LHC searches to be below 144GeV~\cite{CMS_higgs}. 
These limits can be relaxed when the Higgs couplings to gauge bosons is reduced due to the mixing with the singlet. In the parameter space explored, the singlet component is small and the limit is not modified significantly. 
The upper limit can also be relaxed if the Higgs has a large branching fractions into invisible particles. The relic density constraint imposes that light sneutrinos have a mass very near $M_{h_1}/2$. Despite  a phase space suppression factor, the contribution of the invisible mode to the light Higgs decay can in some cases reach 90\% for light sneutrinos, thus relaxing the constraint on the lightest Higgs.  
When imposing  a limit on the light Higgs mass, we have  folded in the effect of the invisible decay modes therefore allowing in a few  cases a light Higgs heavier than 144 GeV.

Observables in the B sector provide powerful constraints on the supersymmetric parameter space assuming minimal flavour violation. The mass differences of B mesons,   $\Delta M_s, \Delta M_d$ have been measured to be
$17.77\pm .12 ps^{-1}$ and $0.507\pm .004 ps^{-1}$~\cite{Nakamura:2010zzi} , somewhat below the SM predictions
\begin{eqnarray}
\Delta M_s^{SM}&=&20.5\pm 3.1 ps^{-1}\nonumber\\
\Delta M_d^{SM}&=&0.59\pm 0.19 ps^{-1}
\end{eqnarray}
Additional supersymmetric contributions of the same sign as the SM ones are therefore strongly constrained, even though there are large uncertainties in the SM prediction mainly due to the CKM matrix elements and hadronic parameters.
Supersymmetric contributions include box diagrams with charged Higgs and quarks, squark/chargino, neutralino  and or gluino loops~\cite{Bertolini:1990if}
as well as  double penguin diagrams with a neutral Higgs exchange. The latter  give a significant contribution for large values of $\tan\beta$ and were not included in our analysis.
The former are important at small values of $\tan\beta$ which are often found in the U(1)' model. In particular  the charged Higgs/quark box diagram contribution adds to the SM contribution inducing too large values for $\Delta M_s$.  The dominant contribution, for small values of $\tan\beta$ is proportional to  $ x \log x (\cot^4\beta)$, where $x=m^2_t/m^2_{H^+}$.  This observable thus constrains severely scenarios with $\tan\beta<1$.
The computation of the mass difference is adapted from the routine provided in  NMSSMTools~\cite{Domingo:2007dx}. We have also used the same estimate for the theoretical uncertainties. 

Other observables such as $b\rightarrow s\gamma, B_S\rightarrow \mu^+\mu^-$ or $B\rightarrow \tau\nu$ are known to receive large supersymmetric contributions when $\tan\beta$ is large, the heavy  Higgs doublet is light and/or the squarks are light. The scenario we will study have heavy squarks,  Higgs doublets above several hundred GeV's and feature  small to moderate values of $\tan\beta$, we therefore have not included these constraints.

\section{Relic abundance of sneutrinos}

\DeclareGraphicsRule{*}{mps}{*}{}
\begin{figure}[!htb]
\begin{center}
\unitlength = 1mm
\begin{fmffile}{1}
\fmfcmd{%
style_def boson_arrow expr p =
cdraw (wiggly p);
shrink (1);
cfill (arrow p);
endshrink;
enddef;}
\begin{fmfgraph*}(60,25)
                              \fmfleft{i1,i2}
                              \fmfright{o1,o2}
                              \fmflabel{$\tilde{\nu}^*_R$}{i1}
                              \fmflabel{$\tilde{\nu}_R$}{i2}
                              \fmflabel{$\bar{b}$}{o1}
                              \fmflabel{$b$}{o2}
                              \fmf{dashes_arrow}{i2,v1,i1}
                              \fmf{fermion}{o1,v2,o2}
                              \fmf{dashes,label=$h_1$}{v1,v2}
                          \end{fmfgraph*} \qquad
\begin{fmfgraph*}(60,25)
                              \fmfleft{i1,i2}
                              \fmfright{o1,o2}
                              \fmflabel{$\tilde{\nu}^*_R$}{i1}
                              \fmflabel{$\tilde{\nu}_R$}{i2}
                              \fmflabel{$W^-$}{o1}
                              \fmflabel{$W^+$}{o2}
                              \fmf{dashes_arrow}{i2,v1,i1}
                              \fmf{boson_arrow}{o1,v2,o2}
                              \fmf{dashes,label=$h_i$}{v1,v2}
                          \end{fmfgraph*}
			  
\bigskip \bigskip \medskip			  			  
			  
\begin{fmfgraph*}(60,25)
                              \fmfleft{i1,i2}
                              \fmfright{o1,o2}
                              \fmflabel{$\tilde{\nu}^*_R$}{i1}
                              \fmflabel{$\tilde{\nu}_R$}{i2}
                             \fmflabel{$\bar{f}$}{o1}
                              \fmflabel{$f$}{o2}
                              \fmf{dashes_arrow}{i2,v1,i1}
                              \fmf{fermion}{o1,v2,o2}
                              \fmf{boson,label=$Z_2$}{v1,v2}
                          \end{fmfgraph*}
\end{fmffile}
\bigskip \medskip
\caption{Main annihilation processes for RH sneutrinos.}
\end{center}
\end{figure}
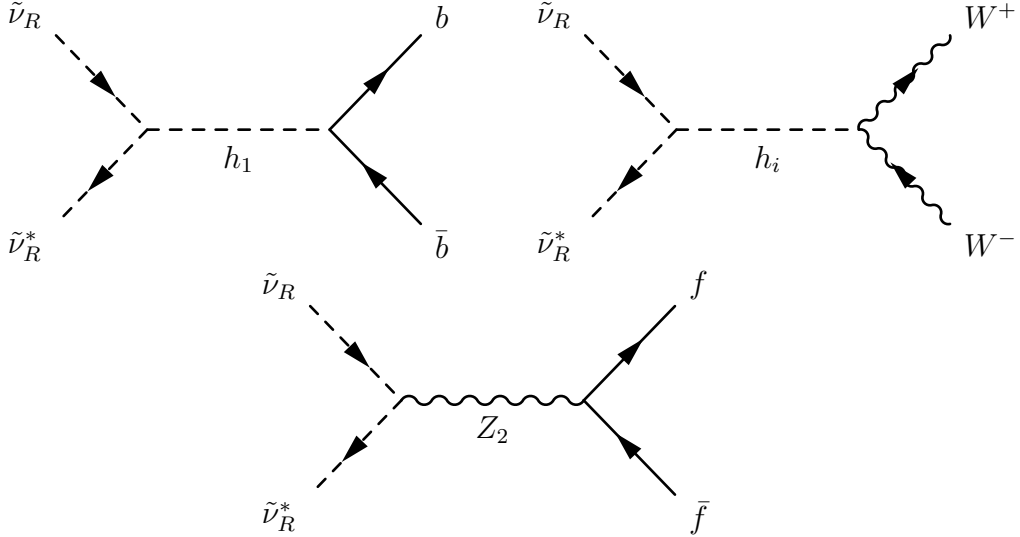

The RH sneutrino couples very weakly to the MSSM particles, its annihilation is therefore primarily through the
particles of the extended sector, the new vector boson, $Z_2$ and new scalars. 
The  coupling of the RHSN to  $Z_2$ is directly proportional to its U(1)' charge, $Q'_\nu$   \begin{equation}
g_{Z_2\tilde{\nu}_{R}\tilde{\nu}^*_{R} }= g_1' Q'_\nu\cos\azz
\end{equation}
This coupling  does not depend much on the other parameters of the model since $\cos\alpha_Z\approx 1$. 
Annihilation of sneutrinos becomes efficient when $m_{\tilde\nu_R}\approx M_{Z_2}/2$  since the s-channel processes can benefit from resonance enhancement.
\newpage
Sneutrino annihilation also become efficient when $m_{\tilde\nu_R}\approx  M_{h_i}/2$ where $h_i$  can be any neutral CP even Higgs field. The couplings of the sneutrino to neutral scalars reads

\begin{equation}
g_{h_i\lsp\lsp} = -{g'_1}^2 Q'_\nu\left[ v \cos\beta Z_{h i1} Q_1  + 
v \sin{ \beta} Z_{h i2} Q_2 +  v_s Z_{h i3} Q'_s \right].\label{eq:hnunu}
\end{equation}
Since $v_s\gg v$, the largest coupling will be to the predominantly singlet Higgs, for which  $Z_{h3i}\approx 1$.Typically the singlet Higgs is the one that has a mass close to $Z_2$, resonant Higgs annihilation will therefore occur for roughly the same sneutrino mass as
the resonant annihilation through $Z_2$. 
The light Higgs is  dominantly doublet, nevertheless its coupling to sneutrinos receives contributions from all three terms in Eq.~\ref{eq:hnunu} since $v_s\gg v$.
This coupling is generally sufficient to have a
large cross section enhancement near $m_{\tilde\nu_R}\approx M_{h_1}/2$. In some cases efficient annihilation can occur away from the
resonance, for example annihilation into W/Z pairs through light Higgs exchange or - for heavier RHSN's - into light Higgs pairs or $t\bar{t}$ pairs through singlet exchange. 
Note that the sneutrino coupling to the lightest Higgs depends on $\lambda$ ($\mu$), an important parameter to determine the mixing of Higgses. 
 For some choice of parameters, the couplings of the mostly doublet $h_1$  to the RHSN can be strongly suppressed, not allowing 
a large enough enhancement on the annihilation cross section.
 Specific examples will be presented in section~\ref{sec:results}.

Sneutrinos can also annihilate in neutrino pairs through t-channel exchange of $\tilde{B}'$ in addition to the usual $Z_2$  contributions. 
This process contributes mostly for light $\tilde{B}'$ and is never the dominant channel.
Finally it is also possible to reduce the abundance of sneutrinos through
coannihilation processes, this occur when the masses of the NLSP and the LSP are within a few GeV's. Coannihilation can
occur with neutralinos, charginos or other sfermions. Typically, because there is only weak couplings of the RH
sneutrino to the rest of the MSSM particles, coannihilation processes involve self-annihilation of the NLSP's and/or NLSP/NNLSP annihilation.
The neutralino NLSP will decay via $\tilde\chi_1\rightarrow \lsp \nu_R$  with  a typical lifetime of  $10^{-19} - 10^{-17} s$ except when it is almost degenerate with the sneutrino LSP which can lead to an increase of lifetime. The NLSP decay will  always occur much before BBN and will not spoil its predictions.

To compute the relic abundance we use \micromegas2.4~\cite{Belanger:2010gh}. For this we first implement the model in LanHEP~\cite{Semenov:2010qt} in both unitary and Feynman gauges and checked gauge invariance for a large number of processes  at  tree-level. 
We then introduced radiative corrections to the Higgs masses in the unitary gauge. This code then produces the model file for \calchep~\cite{Pukhov:2004ca} and \micromegas~\cite{Belanger:2006is}. All facilities of \micromegas~ for computing dark matter properties
are then obtained automatically, in particular the relic density and the LSP-nucleon scattering cross section.  
When describing the numerical results we will always impose the condition that the sneutrino is the LSP. For another study of the neutralino LSP as a
DM candidate see~\cite{Kalinowski:2008iq}.

\section{Direct detection}
\label{sec:psi}

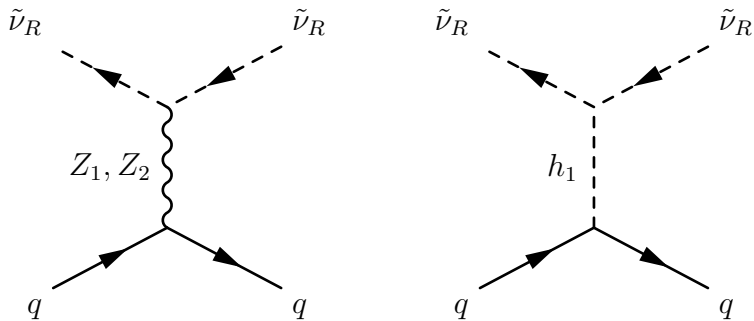
\begin{figure}[!htb]
\begin{center}
\unitlength = 1mm
\begin{fmffile}{2}
\begin{fmfgraph*}(30,40)
			      \fmfbottom{i1,o1}
			      \fmftop{i2,o2}
			      \fmflabel{$q$}{i1}
                              \fmflabel{$\tilde{\nu}_R$}{i2}
                              \fmflabel{$q$}{o1}
                              \fmflabel{$\tilde{\nu}_R$}{o2}
			      \fmf{fermion}{i1,v1}
			      \fmf{fermion}{v1,o1}
			      \fmf{dashes_arrow}{o2,v3}
			      \fmf{dashes_arrow}{v3,i2}
                              \fmf{boson,label=$Z_1,,Z_2$}{v1,v3}
                          \end{fmfgraph*} \qquad \qquad \qquad
\begin{fmfgraph*}(30,40)
			      \fmfbottom{i1,o1}
			      \fmftop{i2,o2}
			      \fmflabel{$q$}{i1}
                              \fmflabel{$\tilde{\nu}_R$}{i2}
                              \fmflabel{$q$}{o1}
                              \fmflabel{$\tilde{\nu}_R$}{o2}
			      \fmf{fermion}{i1,v1}
			      \fmf{fermion}{v1,o1}
			      \fmf{dashes_arrow}{o2,v3}
			      \fmf{dashes_arrow}{v3,i2}
                              \fmf{dashes,label=$h_1$}{v1,v3}
                          \end{fmfgraph*}
\end{fmffile}
\medskip
\caption{Main scattering processes of the RH sneutrino.}
\end{center}
\end{figure}

The cross section for scattering of sneutrinos on nucleons  will be purely spin independent as the sneutrino is a scalar particle. 
 This process receives
contributions from only two types of diagrams: exchange of gauge or Higgs bosons. 
The gauge boson contribution depends on  the vectorial coupling of the fermions
to $Z_{1,2}$, with $Q_V^f= I_3-2Q\sin^2\theta_W$ for $Z^0$ and $Q_V'^f=Q'_f-Q'_{f^c}$ for Z'.
 The total gauge boson contribution to the spin independent cross section on point-like nuclei reads
\begin{eqnarray}
\sigma^{\rm SI,\,Z}_{\lsp N} =  \frac{\mu_{\tilde\nu N}^2}{\pi} (g'_1 Q'_\nu)^2  
\left[\left(y(1-4\sin^2\theta_W)+ y'\right) Z + (-y+ 2y')(A-Z)\right]^2
\label{eq:ddz}
\end{eqnarray}
where 
\begin{eqnarray}
y &= &\frac{g_1}{4} \sin\alpha_Z\cos\alpha_Z\left(\frac{1}{M_{Z_2}^2}- \frac{1}{M_{Z_1}^2}\right)\nonumber\\
y'&= & - \frac{g'_1}{2} Q_V'^d \left(\frac{\sin^2\alpha_Z}{M_{Z_1}^2} + \frac{\cos^2\alpha_Z}{M_{Z_2}^2}\right)
\end{eqnarray}
and $Q_V'^d=-4/\sqrt{40} \cos\tesix$.
The contribution in $1/M^2_{Z_2}$ in $y$ and the one proportional to $\sin^2\azz$ in $y'$ are always suppressed.  
 From
Table~\ref{tab:u1} one sees that all fermions have a purely axial-vector couplings to   $Z_\psi$  and that  u-quarks also 
have purely axial vector couplings to $Z_\chi$. Therefore the Z' contribution is  solely dependent on its coupling to d-quarks,
hence the term in $Q_V'^d$ in $y'$. This contribution  
is expected to be twice as large for neutrons than for protons, see Eq.~\ref{eq:ddz}. 
In the model $U(1)_\psi$, $y'=0$ and
the cross section will depend on the $Z_1$ exchange contribution.  In this case the amplitude for protons is suppressed by a 
factor $1-4\sin^2\theta_W$ as compared to that for neutrons. Furthermore the $Z_1$ contribution proceeds through the $Z'$ component 
so that the cross section is proportional to $\sin^2\alpha_Z$. 
When $\cos\tesix \neq 0$  the term in y' usually dominates by about one order of magnitude for a TeV scale $Z_2$ and the mixing angle
$\azz=10^{-3}$. In these scenarios only a  weak dependence on $\alpha_Z$ is expected,
the largest cross sections are expected for $\tesix\simeq 0$ corresponding to the maximal value of $Q_V'^d$.

The Higgs contribution leads to a cross section 
\begin{equation}
\sigma^{{\rm SI,}\,h}_{\lsp N} =\frac{\mu_{\tilde\nu N}^2 }{4\pi}  \sum_i \frac{g_{{h_i\lsp\lsp}}^2 } {m_{h_i}^4 m_{\lsp}^2} 
\left( (A-Z) \sum_q g_{{h_i}qq} f_q^n m_n + Z \sum_q g_{{h_i}qq} f_q^p m_p   \right)^2  \,,
\label{eq:ddh}
\end{equation}
where $g_{{h_i}qq}=-eZ_{hi1}M_q/(2 M_W s_W c_\beta)$ is the Higgs coupling to quarks after the quark mass
has been factored out and $g_{h_i\lsp\lsp}$ is defined in Eq.~\ref{eq:hnunu}. 

Because of the dependence on the Higgs mass and the fact that the Higgs couplings to quarks goes only through the doublet component, 
the lightest Higgs doublet generally gives the dominant contribution. 
Note that the Higgs contribution is inversely proportional to the square of the sneutrino mass and is therefore expected to dominate 
at low masses since the Z contribution depends only weakly on the sneutrino mass through the effective mass, $\mu_{\tilde\nu N}$. Furthermore the Higgs contribution is roughly the same for neutrons and protons.
The quark coefficients of the nucleon were taken to be the default values in micrOMEGAS2.4 
(with $f^p_{u,d,s}= 0.033,0.023,0.26 \; {\rm and} \; f^n_{u,d,s}=0.042,0018,0.26$)~\cite{Belanger:2008sj}.
There  can be large uncertainties in these coefficients, in particular recent lattice results point towards a smaller s-quark coefficient~\cite{Giedt:2009mr}, however the quark coefficients  will have a significant impact only when the Higgs contribution is dominant, that is for  sneutrinos below $\approx 100$~GeV.

The total spin independent cross section on point-like nucleus is 
obtained after averaging over the sneutrino and anti-sneutrino. Note that the interference between the Z and H exchange diagrams have
opposite signs for $\tilde{\nu}N$ and $\tilde{\nu}^*N$. 
Here we assume equal numbers of sneutrino and anti-sneutrinos so that the total cross section is the sum of the Z and H contributions.
To compute the direct detection cross section we use \micromegas2.4 and thus include all the contributions.
To take into account the fact that the proton and neutron contribution are not necessarily equal, 
we compute  the normalized cross-section on a point-like nucleus,  
\begin{equation}
\sigma^{\rm SI}_{\lsp N}= \frac{4 \mu_\chi^2}{\pi}\frac{\left( Z f_p+ (A-Z)f_n\right)^2}{A^2 }
\end{equation}
where the average over $\lsp$ and $\lsp^*$ is assumed implicitly. 
This cross-section can be directly compared with the limits on $\sigma^{\rm SI}_{\tilde\chi p}$ that is extracted from each
experiment.

\section{Results}
\label{sec:results}
The free parameters of the model are
$\mlsp, \mu, M_1,M_2,M'_1,M_K,A_\lambda$, $M_{Z_2},\theta_{E_6},\azz$,  as well as all masses and trilinear couplings 
of sleptons and squarks. 
To reduce the number of free parameters we fix those that do not belong to the sneutrino or neutralino/Higgs sectors. 
We thus fix the soft masses of sleptons and squarks to 2 TeV and we take $A_f=0$ and $A_t=1$TeV. We furthermore
assume  $M_1=M_2/2=M_3/6$ as dictated by universality at the GUT scale and  $M_K=1$~eV since this parameter 
mainly affects the $\bino-\binop$ mixing and is not directly relevant for our study. 
The remaining  free parameters are $\mlsp,\mu,M_1,M_1',A_\lambda$ and  $M_{Z_2},\azz$. We first consider specific choices of $\theta_{E_6}$ before letting it be a free parameter. 

\subsection{The case of the $U(1)_\psi$ model}
\label{sec:psi}
As we have discussed above, we expect the relic abundance of the sneutrino dark matter to be generally too high.
The only ways to bring the abundance within the range preferred by WMAP will be to enhance the annihilation through a resonance effect or
make use of coannihilation. We will first describe the typical behaviour of $\Omega h^2$ as a function of the $\lsp$ mass
for fixed values of the free parameters. Once $\tesix$ is fixed the parameters that have a strong influence on dark matter are
$M_{Z_2}$, the mass of the new gauge boson,
$A_\lambda$  that influence the Higgs spectrum, as well as
$\mu,M_1$ that determine the region where the sneutrino is LSP through their influence on the neutralino mass.
The parameter $\mu$  also influences the Higgs mixing matrix and therefore the coupling of the Higgses to sneutrinos.   
To limit the number of free parameters we will first fix $M_1=M_1'=1$~TeV and rather modify the property of the neutralino NLSP by varying $\mu$. 
Furthermore allowing to have   $\mu<M_1$ will cover the case of the higgsino NLSP which has much higher annihilation cross section 
than the bino and is  therefore more likely to  play an important role in coannihilation channels.

\subsubsection{A case study  with $M_{Z_2}=1.6$~TeV}
\label{sec:psi:mz}
We first consider the case  where the new gauge boson is just above the LHC exclusion limit, that is $M_{Z_2}=1.6$~TeV and we fix  $\mu=1$~TeV,  $\azz=-0.001$ and $A_\lambda=1.5$~TeV. 
For this choice of parameters, the mass spectrum is such that $h_2$ is dominantly singlet and has roughly the same mass as the $Z_2$ while  
the other heavy Higgses ($h_3,A^0,h^+$) are nearly degenerate with a mass around 1.97TeV. The  light Higgs has a mass of 136.6~GeV and is safely above the LEP
limit despite the small value of $\tan\beta=2.05$.  The lightest neutralino has a mass of 957~GeV and is a  mixture of bino and higgsino.
The sneutrino is therefore the DM candidate when its mass ranges from a few GeV's up to the mass of the lightest neutralino.
The value of  $\Omega h^2$ lies below the WMAP upper bound in two regions, the first when  $60 {\rm GeV}\le m_{\lsp} \le 69 {\rm GeV}$
the second when $734{\rm GeV}\le m_{\lsp} \le 805{\rm GeV}$.
In both regions the annihilation cross section is enhanced significantly by a resonance effect. 
In the first region the $h_1$ exchange is enhanced and the preferred annihilation channel is into $b{\bar b}$ pairs while in the second region
the $h_2$ and  $Z_2$ exchange are enhanced  with a dominant contribution from $h_2$.  
The preferred annihilation channels are into $W^+W^-,t\bar{t},h_1h_1,Z_1Z_1$ pairs,  the dominant decay modes of $h_2$. 
Note that the decays into gauge bosons proceed through the small doublet component of $h_2$. 
For larger  masses of the sneutrino, coannihilation with the lightest neutralino can take place. 
Coannihilation processes involving pairs of neutralinos  annihilating through the heavy pseudoscalar Higgs 
 can decrease $\Omega h^2$ such that  the WMAP upper bound is satisfied.  
This occurs when the NLSP-LSP mass difference drops below 60 GeV.

 \begin{figure}[!ht]
\includegraphics[width=8cm,height=6.5cm]{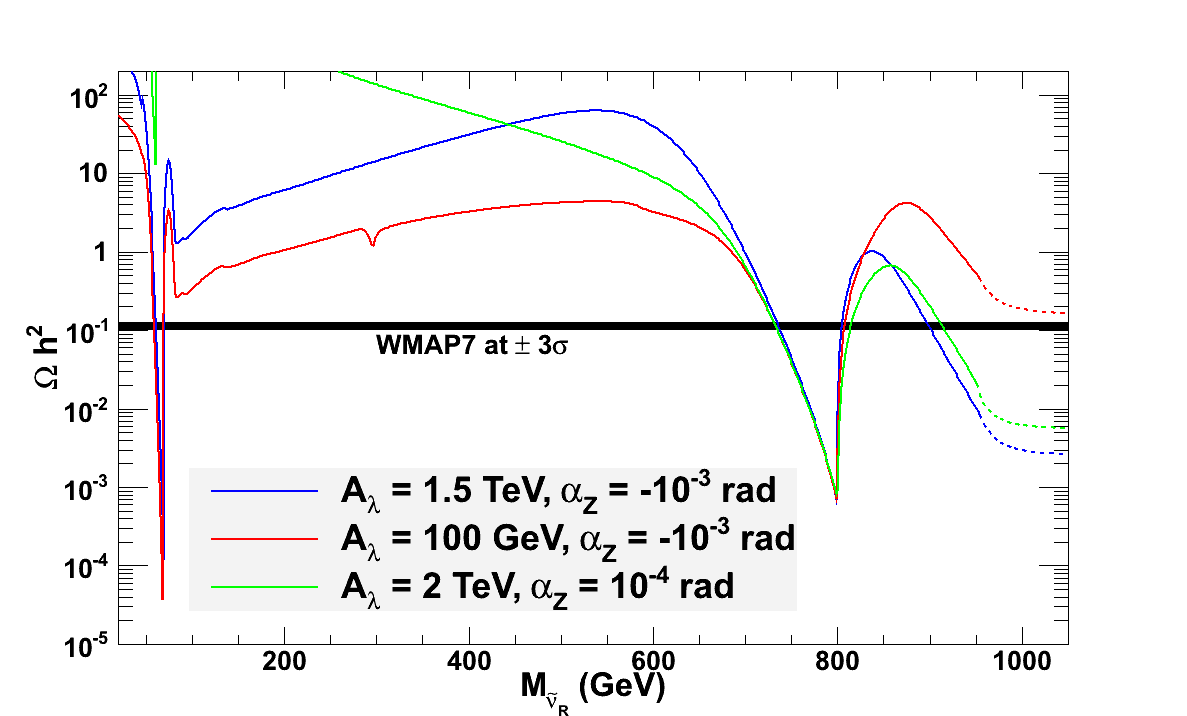} 
\includegraphics[width=8cm,height=6.5cm]{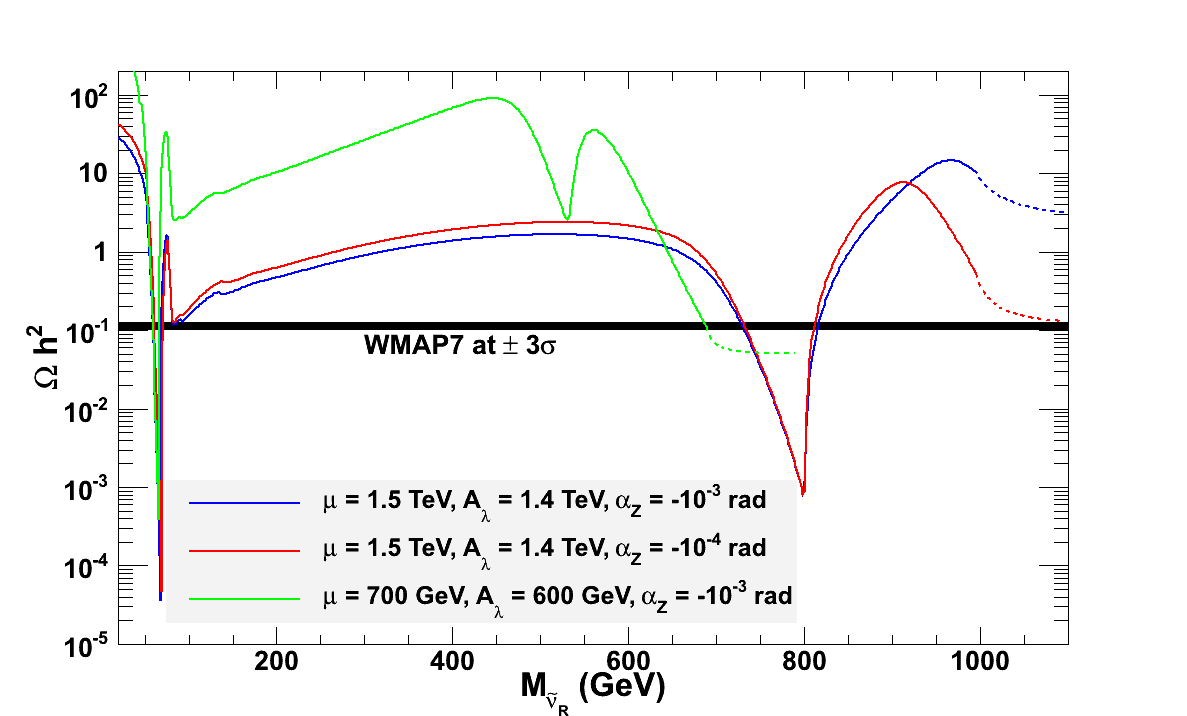} 
\ablabels{63}{144}{58} 
   \caption{$\Omega h^2$ as a function of the LSP mass for
    $\mzp=1.6$~TeV and (a) $\mu=1$~TeV,  $(A_\lambda(TeV), \azz)=(1.5,-10^{-3}),(0.1,-10^{-3}),(2.,-10^{-4})$
    (b)  $\mu=1.5{\rm TeV}$, $(A_\lambda(TeV), \azz)=(1.4,-10^{-3}),(1.4,-10^{-4})$ and 
    $\mu=0.7{\rm TeV}$,  $(A_\lambda(TeV), \azz)=(0.6,-10^{-3})$. Full (dash) lines correspond to the region where the LSP is the RHSN (neutralino). }
\label{fig:omega}
\end{figure}

This general behavior is somewhat influenced by our choice of $A_\lambda$,  $\mu$  and $\alpha_Z$ since
these parameters influence the masses and mixings of the Higgs doublets and the couplings of sneutrinos to Higgses.
For example for $A_\lambda=100$~GeV, the mass of the heavy doublet (which is now $h_2$) decreases to 592 GeV.
The $h_2/A^{0}$ exchange contributes to sneutrino annihilation into top pairs  thus leading to a decrease of the  value of $\Omega h^2$ as one approaches the Higgs resonance,
the drop is not significant enough to bring the relic density below the WMAP upper bound, see Fig.~\ref{fig:omega}a. 
As another example consider the case  $A_\lambda=2$~TeV and $\alpha_Z=10^{-4}$. Here the $h_1\lsp\lsp$ coupling is suppressed,
$\Omega h^2$ becomes very large and despite a resonance effect when  $m_{\lsp} \approx m_{h_1}/2$ the
WMAP upper bound can never be  satisfied for a sneutrino lighter than 100GeV, see Fig.~\ref{fig:omega}a. 

The parameter $\mu$ induces corrections to the light Higgs mass as well as shifts in the Higgs mixing matrix. In particular increasing $\mu$ 
(and therefore $\lambda$) to 1.5~TeV  increases the singlet mixing in the light Higgs and thus the $\lsp\lsp h_1$ coupling. 
This makes  annihilation processes through Higgs exchange more efficient, and gives rise to 
a new region where $\Omega h^2$ is below the WMAP upper bound when the sneutrino mass is  just above the W pair threshold, see Fig.~\ref{fig:omega}b with $A_\lambda=1.4$~TeV. Note that in this case the lightest  neutralino has a large bino component and coannihilation is not very efficient.
Lowering $\mu$ to  700 GeV, changes the nature of the lightest neutralino which becomes dominantly higgsino with  its mass determined by $\mu$. Thus  the range of masses where the sneutrino is the LSP becomes narrower. In fact the singlet Higgs/$Z_2$ pole annihilation region cannot be reached  when the sneutrino is the LSP, the region compatible with WMAP is rather one where higgsino coannihilation dominates. In Fig.~\ref{fig:omega}b, one can see a significant drop in $\Omega h^2$ 
near the $h_2$ resonance, this is however not sufficient to have efficient annihilation of the RHSN.
 
\begin{figure}[!htb]
\includegraphics[width=8cm,height=6.5cm]{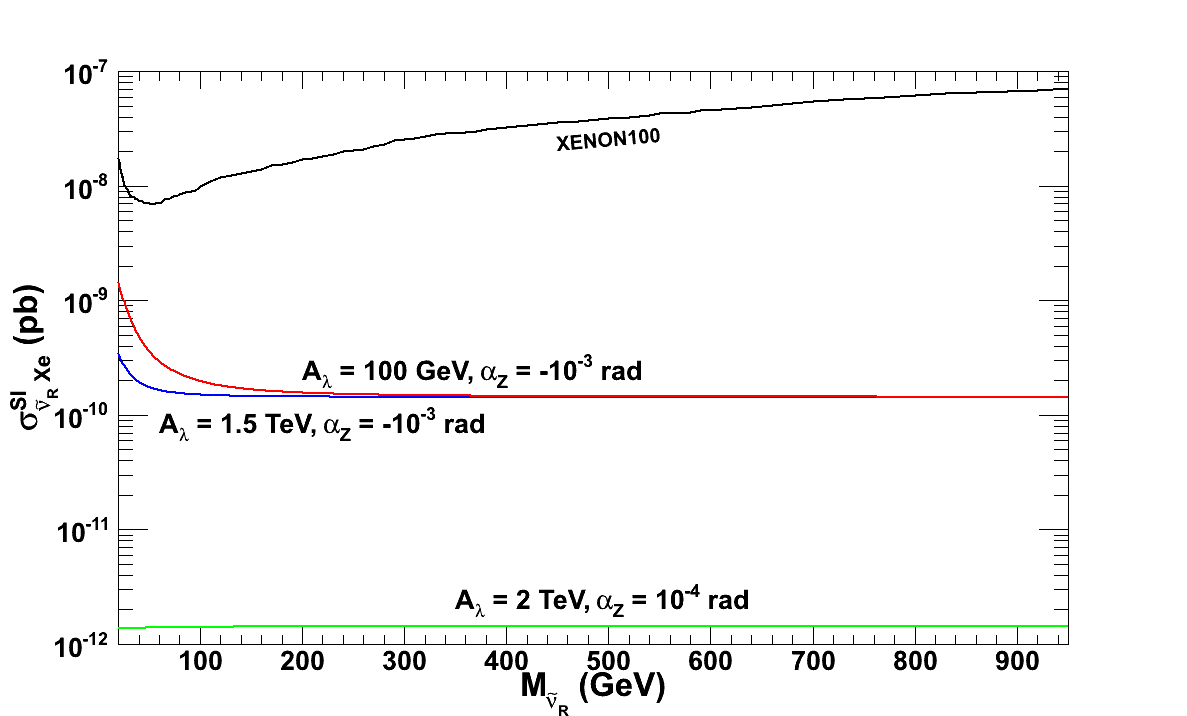} 
\includegraphics[width=8cm,height=6.5cm]{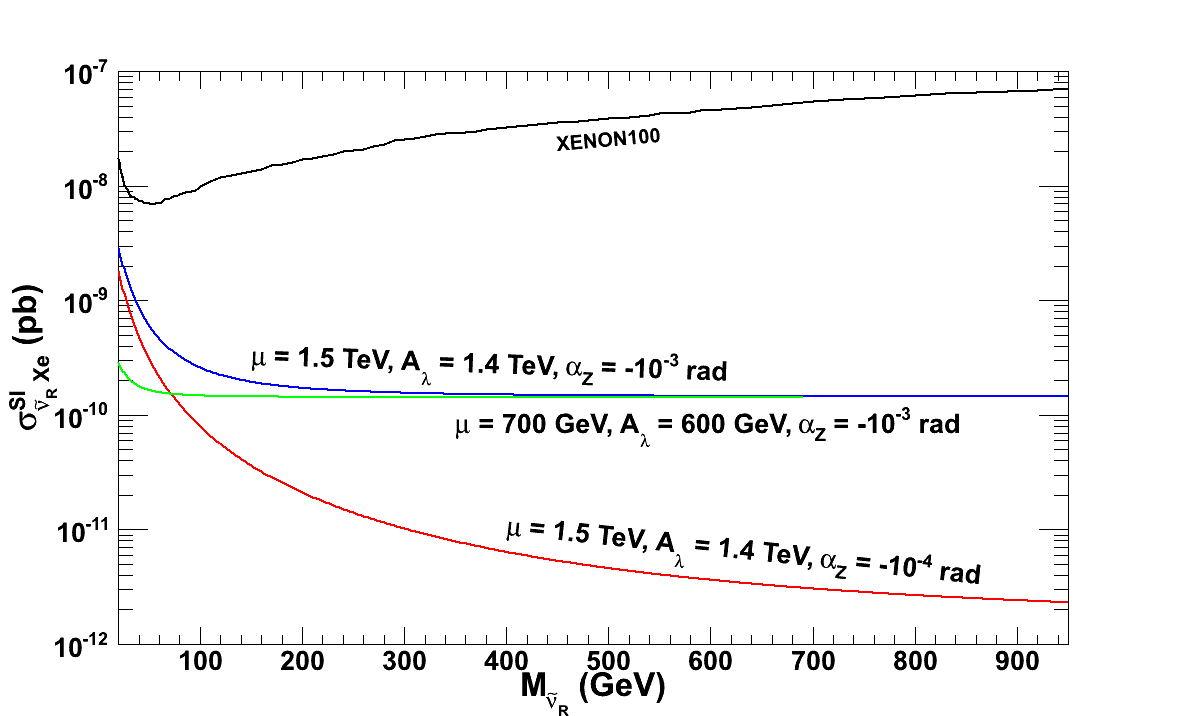} 
\ablabels{11}{92}{58} 
   \caption{$\sigma^{SI}_{\lsp N}$ as a function of the LSP mass for the same parameters as Fig.~\ref{fig:omega}.
 The Xenon100 exclusion curve is also displayed.}
\label{fig:dd}
\end{figure}
 
The direct detection spin independent (SI) cross section receives contributions from both the light Higgs and $Z_1$ exchange.
In general for the $U(1)_\psi$ scenario, the cross-section  is  below the limits from Xenon100~\cite{Aprile:2010um}. As we have argued above, this is because the 
$Z_1$ contribution is directly proportional to $\sin^2\azz$. This contribution nevertheless dominates  for  a mixing angle $|\azz|=10^{-3}$  except for small masses.  It gives  $\sigma  =1.4\times 10^{-10}$~pb for $\mlsp \geq 200$~GeV and is mostly independent of other input parameters, see Fig.~\ref{fig:dd}a,b.
Furthermore the $Z_1$ contribution  is much larger for neutrons than for protons. We take that into account by computing the cross section for scattering on point-like nucleons (here we use Xenon) normalized to one nucleon. 
For masses in the range $50-100$~GeV, the SI cross section is enhanced due to the light Higgs exchange contribution which increases at low sneutrino masses, although for this scenario the values are always below the experimental limit.
The predictions  are in the range $\sigma=2\times 10^{-10}- 3\times 10^{-10}$~pb for $\alpha=-10^{-3}$.
The SI cross section drops rapidly with the mixing angle, for example for $\azz=|10^{-4}|$, $\ssi \simeq 10^{-12}$~pb for $\mlsp \geq 200$~GeV, see
Fig.~\ref{fig:dd}a. For small values of the  mixing angle, the $Z_1$ contribution is suppressed and the Higgs contribution can become dominant even for masses of a few hundred GeV's, this however corresponds to a low overall cross section, see for example Fig.~\ref{fig:dd}b. 
The spin-independent cross section can be further suppressed in the limit $\azz=0$ and for parameters for which  the $h_1$ couplings to sneutrino is suppressed.

\subsubsection{Exploration of $U(1)_\psi$ parameter space}

After having describe the general behavior of $\Omega h^2$ and of the direct detection rates for some choice of parameters, 
we now search for the region in parameter space that are compatible with 
$0.1018 < \Omega h^2 <0.1228$ corresponding to the WMAP  $3\sigma$ range~\cite{Komatsu:2010fb} as well as with the direct detection
limit of Xenon100~\cite{Aprile:2010um}. We have further imposed 
the limits on the Higgs sector from LEP and the LHC and have taken into account the impact of the invisible decay mode of the Higgs.
We have also used limits on sparticles from LEP as implemented in \micro~. 
Note  that since we are only considering the case of heavy sfermions, these limits as well as  the new LHC limits on sparticle masses do not come into play
~\cite{Aad:2011ib,Collaboration:2011zy}.   Finally we have also imposed the constraints from $\Delta M_{d,s}$.
We have performed random scans with $5\times 10^6$ points. The range for the parameters are listed in Table~\ref{tab:e6}. 
 As before we here fix all sfermion masses to 2TeV and neglect all  trilinear couplings with the exception of $A_t=1$~TeV.
  Note that we impose $\mu>0$ and therefore consider only positive values for $A_\lambda$, see Eq.~\ref{eq:ma}.

\begin{table}[!htb]
\begin{center}
\begin{tabular}{|c|c|}\hline
Parameter & Range \\ \hline \hline
$\mlsp$ & [0, 1500] GeV\\ 
$M_{Z_2}$ & [1300, 3000] GeV\\ 
$\mu$ & [100, 2000] GeV\\ 
$A_\lambda$ & [0, 2000] GeV\\
$\azz$ & [-0.003, 0.003] rad \\ 
$M_1$, $M'_1$ & [100, 2000] GeV\\ \hline
\end{tabular}
\caption{Range of the free parameters in the $U(1)'$ models \label{tab:e6}}
\label{tab:range}
\end{center}
\end{table}

\begin{figure}[!htb]
\includegraphics[width=8cm,height=6.5cm]{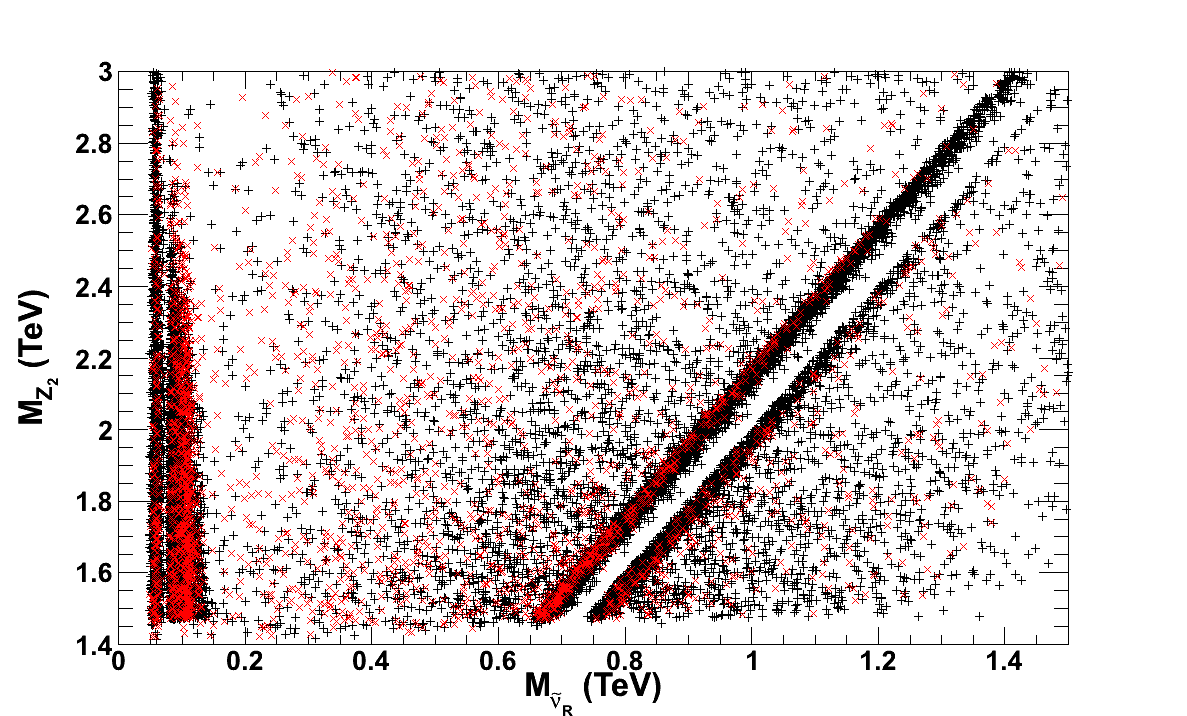} 
\includegraphics[width=8cm,height=6.5cm]{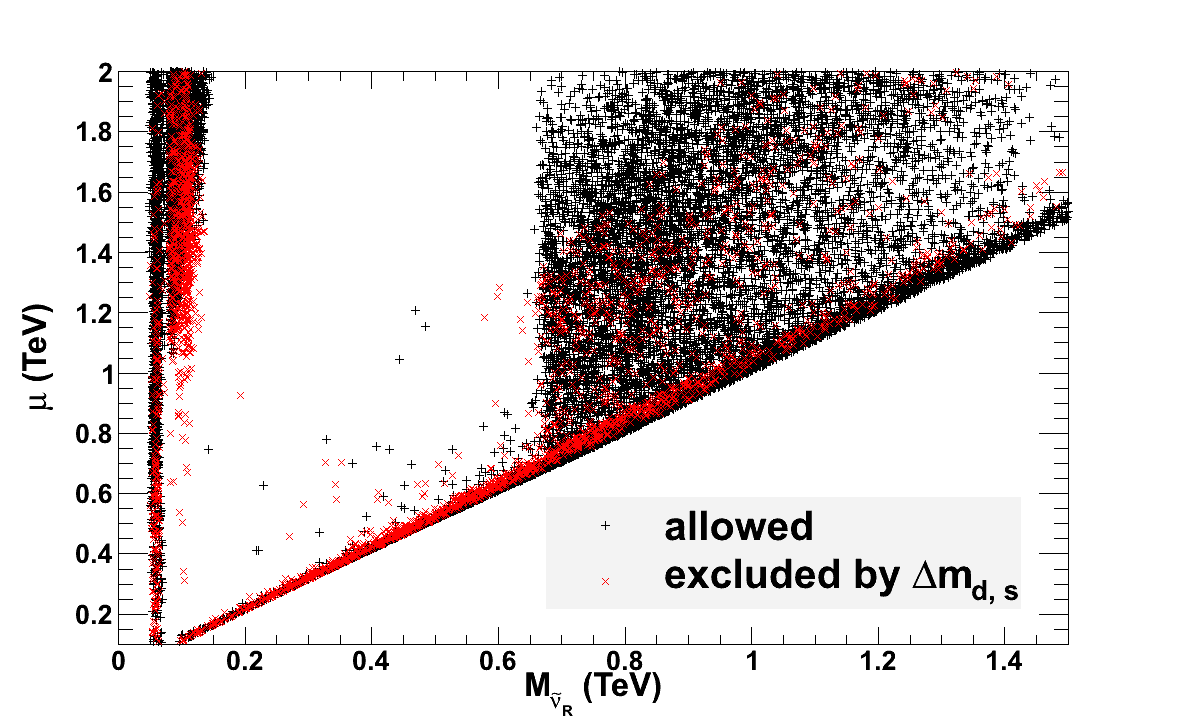} 
\ablabels{15}{96}{58}
  \caption{The allowed scenarios in the   a) $M_{Z_2}$ vs $m_{\lsp}$ plane and 
    b)  $\mu$ vs $m_{\lsp}$ plane. Points excluded by the $\Delta M_{d,s}$ constraint are displayed in red.}
\label{fig:random_psi}
\end{figure}

The results of the scan  as displayed in Fig.~\ref{fig:random_psi} show that 
the allowed values for $m_{\lsp}$ cover a wide range from 55~GeV  to the largest value probed.  
The allowed points in the plane $m_{\lsp}- M_{Z_2}$ are clustered in three regions around $m_{\lsp}\approx 60$~GeV, around $m_{\lsp} \approx M_{Z_2} -\delta (+\delta')$ with $\delta \approx 70$~GeV$(\delta' \approx 10$~GeV), and $m_{\lsp}\approx 90-100$~GeV. The first two  are characterized by  the main annihilation mechanism near a resonance. The latter corresponds to annihilation through a Higgs exchange. As discussed above, this requires a large value for $\mu$, see Fig.~\ref{fig:random_psi}b.
The remaining allowed scenarios correspond to $m_{\lsp} \approx m_{\chi^0_1}$. In most cases the NLSP is a higgsino and $m_{\lsp}\approx \mu$, see Fig.~\ref{fig:random_psi}b, then 
important contributions from coannihilation processes involving  neutral and charged higgsinos annihilating into fermion pairs
are to be added  to  the sneutrino annihilation processes dominated by the channels  into $WW,ZZ$ through Higgs exchange. 
Because  $\mu$ is  constrained by the LEP limit on charginos, the RHSN mass is in this case heavier than 90~GeV.
A few cases where  the NLSP has an important bino component  are also found, those correspond to  the points  above the line $\mu \approx \mlsp$ in  Fig.~\ref{fig:random_psi}b.

In this model, the cross section for LSP scattering on nuclei is dominated by either Higgs or $Z_1$ exchange and is mostly independent
of $M_{Z_2}$.
The direct detection cross section  can reach at most $\ssi= 2 \times 10^{-9}$~pb when $m_{\lsp}\approx 50-60$~GeV  and the light Higgs exchange
dominates while the maximal value is only $\ssi \approx 5\times 10^{-10}$~pb for sneutrinos above 200~GeV when $Z_1$ exchange is dominant.
These predictions are much below the exclusion limits of XENON100~\cite{Aprile:2010um}.
Values below $\ssi= 10^{-13}$~pb can also be obtained, these occur 
when the $Z_1$ contribution is suppressed by the small mixing angle and the light Higgs coupling to the LSP are suppressed as well.
Cases where coannihilation processes dominate can also lead to small cross sections.

\begin{figure}[!htb]
\includegraphics[width=8cm,height=6.5cm]{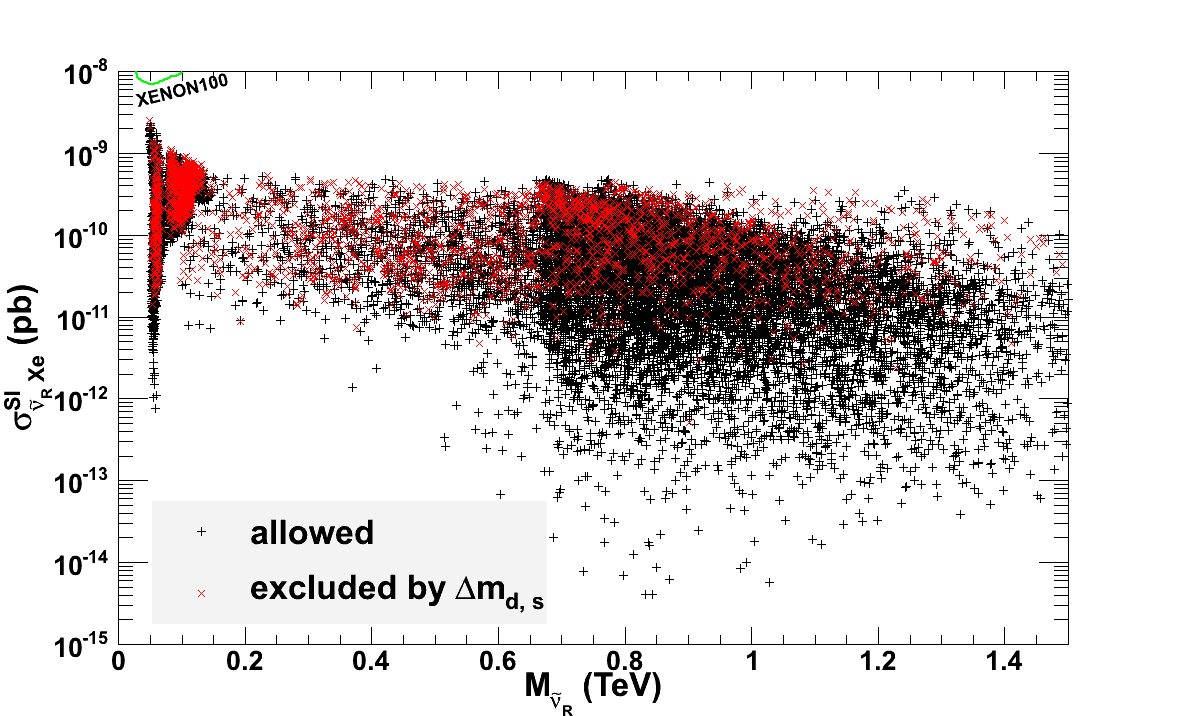} 
\includegraphics[width=8cm,height=6.5cm]{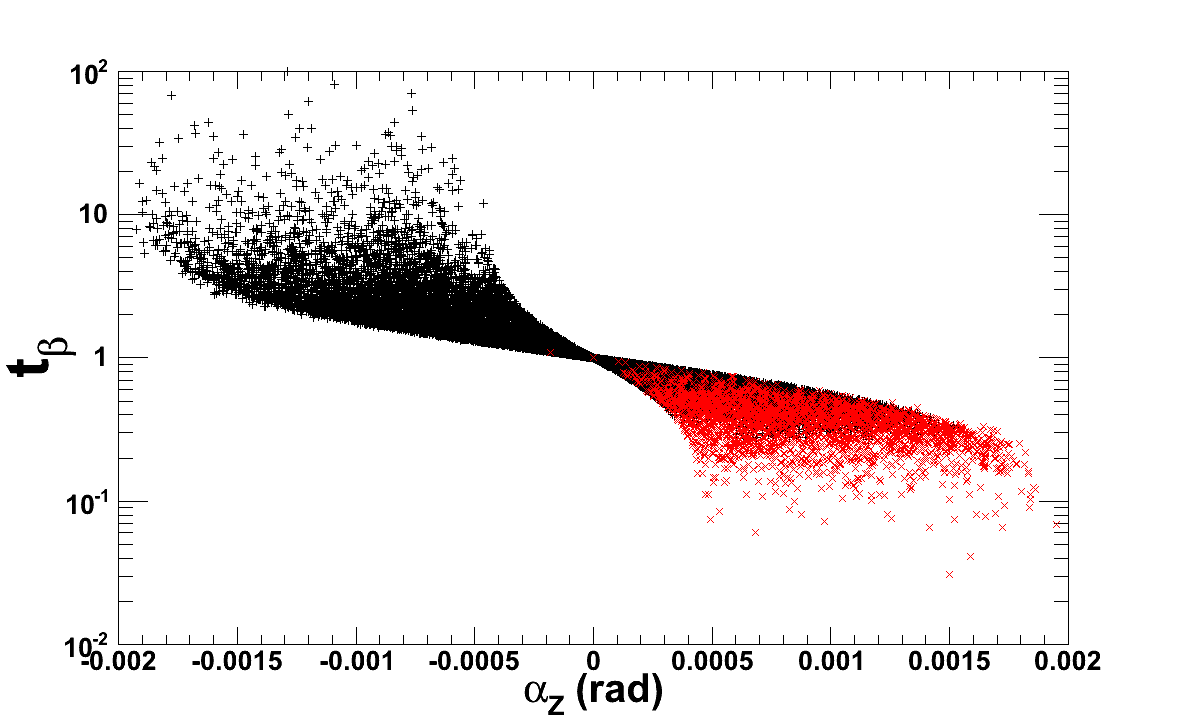}
\ablabels{63}{144}{58} 
\caption{(a)
$\sigma^{SI}_{\lsp Xe}$ as a function of the LSP mass for the allowed scenarios in the $U(1)_\psi$ model.
(b) Allowed scenarios in the $\tan\beta$- $\mlsp$ plane. Points excluded by $\Delta M_s$ are displayed in red.  }
\label{fig:random_psi_direct}
\end{figure}

Constraints from $\Delta M_{d,s}$ rule out some of the parameter space, in particular the case where $\alpha_Z>0$ since it leads to values of $\tan\beta<1$.
The allowed points in the $\tan\beta -\alpha_Z$ plane are displayed in Fig.~\ref{fig:random_psi_direct}b. 
As mentioned in Section~\ref{sec:constraint}, when $\tan\beta$ is small the charged Higgs box diagram adds to the SM contribution and leads to  value for $\Delta M_s$ that is too large. This constrain  does not influence directly the range of  predictions for the direct detection cross section, nor does it affect the range of allowed masses for the RHSN. 

\subsection{The case of the $U(1)_\eta$ model}

The properties of the RHSN dark matter are  dependent on the choice of the U(1)' charges.
To illustrate some of the differences we present results for 
$\tesix=-.29\pi$ ($U(1)_\eta$) before discussing arbitrary values in the next section. 
We chose this  model  for its phenomenological properties. 
For this value of $\tesix$, the coupling of the sneutrino to $Z_2$ is 
enhanced as compared to the previous example with $Q'_\nu=-\sqrt{5}/2\sqrt{3}$. Furthermore the coupling of the RHSN to 
$h_1$ is typically enhanced as compared to the model $U(1)_\psi$. 
 Finally  the vectorial couplings of $Z'$ to neutrinos and d-quarks are non-zero. 
Therefore both terms in Eq.~\ref{eq:ddz} can contribute to the SI
cross section, this implies  a direct detection cross section that is much larger than found previously.
Nevertheless the vectorial couplings of $Z_1$ are  far enough from their maximal value that it is possible to find scenarios that satisfy direct
detection limits as will be demonstrated below.

 \subsubsection{A case study with $M_{Z_2}=1.6$~TeV}

\begin{figure}[!htb]
\includegraphics[width=8cm,height=6.5cm]{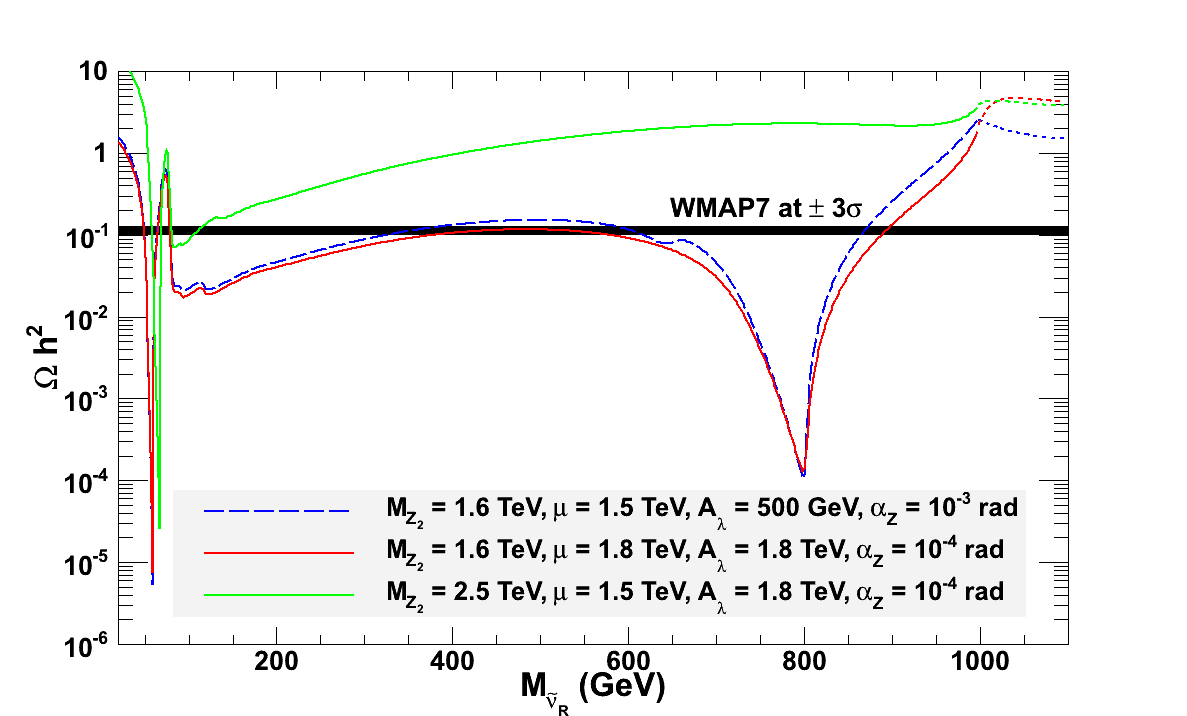} 
\includegraphics[width=8cm,height=6.5cm]{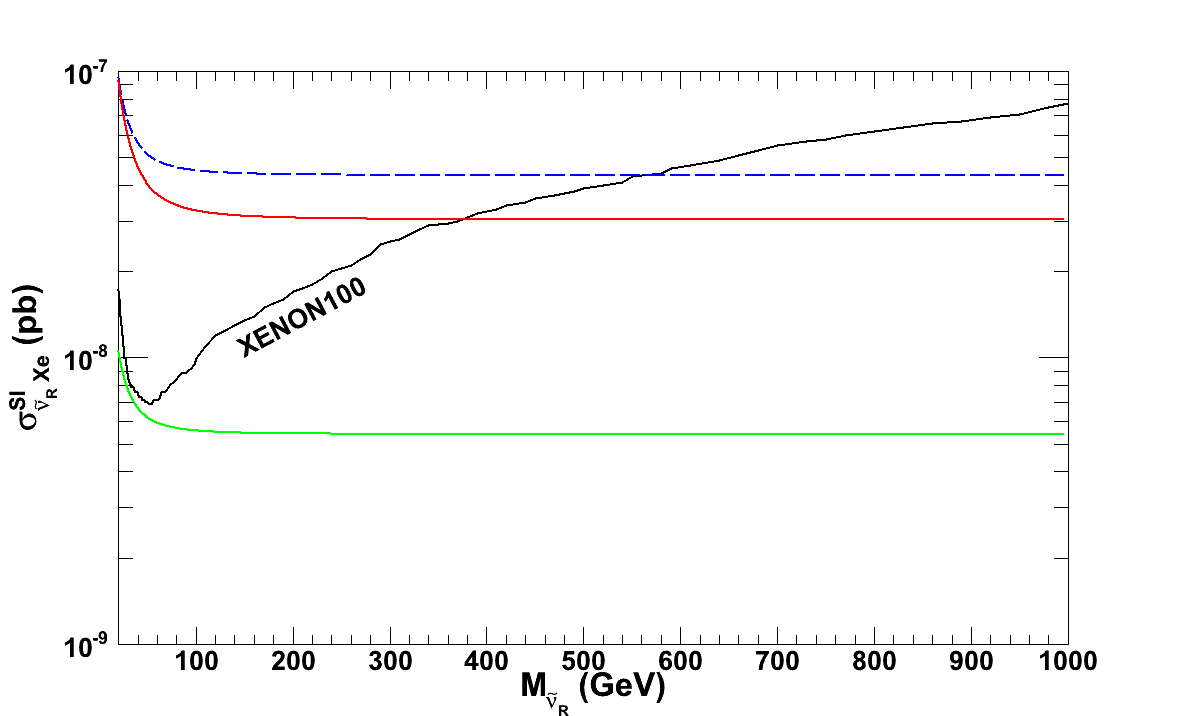}
\ablabels{12}{92}{58}
   \caption{$\Omega h^2$ as a function of the LSP mass for a)  $\mzp=1.6$~TeV,
 $\mu=1.5$~TeV, $A_\lambda=0.5$~TeV, $\azz=10^{-3}$ and 
 $\mzp=1.6,2.5$~TeV, $\mu=1.8$~TeV, $A_\lambda=1.8$~TeV, $\azz=10^{-4}$ . The dotted lines correspond to the
 neutralino LSP.
b) $\ssi$ as a function of the LSP mass for the same choice of parameters.}
\label{fig:omega_eta}
\end{figure}

To illustrate the behaviour of the relic density and the direct detection rate we choose $M_{Z_2}=1.6$~TeV, $\alpha_Z=0.001$,
 $\mu=1.5$~TeV and $A_\lambda=0.5$~TeV. For this choice of parameters $m_{h_1}=116.9$~GeV despite a small value of $\tan\beta=1.2$. The doublet Higgses are around  $1.3$~TeV while $h_3$ is nearly degenerate with $Z_2$. 
As for the the model $U(1)_{\psi}$, the relic density satisfies the WMAP upper bound in the regions where annihilation into light Higgs
or singlet Higgs/$Z_2$ is enhanced by a resonance effect as well as in a  region where annihilation 
via $h_1$ exchange is efficient without the benefit of a
resonance enhancement.  The latter region extends over a wide range of values for the LSP mass because of the large couplings of the sneutrinos to $h_1$.
The preferred annihilation channels are typically into  WW, ZZ $t\bar{t}$ or $h_1h_1$ pairs as well as  into
$b\bar{b}$ for light sneutrinos. Note also that exchange of $h_2$ can contribute significantly to sneutrino annihilation, see the small dip at $\mlsp=650$~GeV in Fig.~\ref{fig:omega_eta}a. 
In this scenario  the direct detection cross section is large ($\ssi\approx 4.5\times 10^{-8}$~pb for $\mlsp \geq 100 {\rm GeV}$) and exceeds the Xenon100 bound for a sneutrino LSP below 550 GeV, see Fig.~\ref{fig:omega_eta}b. This is because  both the $Z_1$ and $Z_2$ contribute significantly to the cross section. 
Extending the region of validity of sneutrino DM over a larger mass range after considering direct detection limits thus requires decreasing the mixing angle $\alpha_Z$ and/or  increasing the mass of the $Z_2$. 
For example taking $\alpha_Z=10^{-4}$, $\mu=A_\lambda=1.8$TeV has the effect of decreasing the relic density - so that it is below the WMAP upper bound for sneutrino masses in the range 90-900~GeV -  while also  decreasing the SI cross section. Yet the LSP is still constrained to be $\mlsp \geq 370$~GeV. In fact  the direct detection rate decreases by less than a factor of two, this is because  the contribution of the $Z_2$ exchange, the term proportional to 
$Q'^d_V$ in Eq.~\ref{eq:ddz}, is not suppressed by $\sin\alpha_Z$.
This means that to have sneutrino DM with a mass of a few hundred GeV's  also requires increasing the mass of the $Z_2$. For example 
taking $M_{Z_2}=2.5$~TeV decreases the direct detection rate below the Xenon100 exclusion for all masses of the LSP. 
However for this choice of parameters,  only light sneutrinos also satisfy the relic density constraint since the heavy sneutrino LSP 
cannot annihilate through the singlet/$Z_2$ resonance.

Finally we comment on the coannihilation region. We have fixed  $M_1=1$~TeV, so  that the lightest neutralino has a mass  $\approx 1$~TeV and is dominantly bino. For values of $\mu \leq 1$~TeV, the lightest neutralino  has a large higgsino component and coannihilation is efficient. For the benchmarks with $\mu=1.5$~TeV
the coannihilation processes involving binos are  not efficient enough to reduce $\Omega h^2$ to a value compatible with WMAP.

 \subsubsection{Exploration of $U(1)_\eta$ parameter space}
 
\begin{figure}[!htb]
\includegraphics[width=8cm,height=6.5cm]{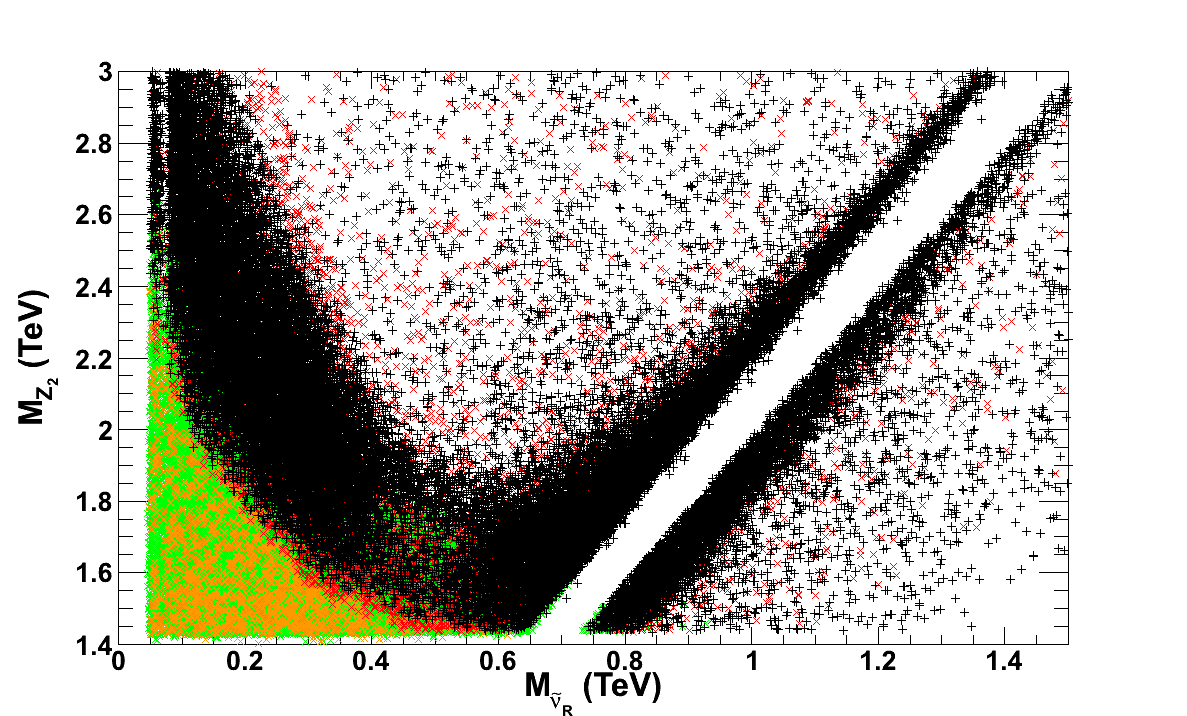} 
\includegraphics[width=8cm,height=6.5cm]{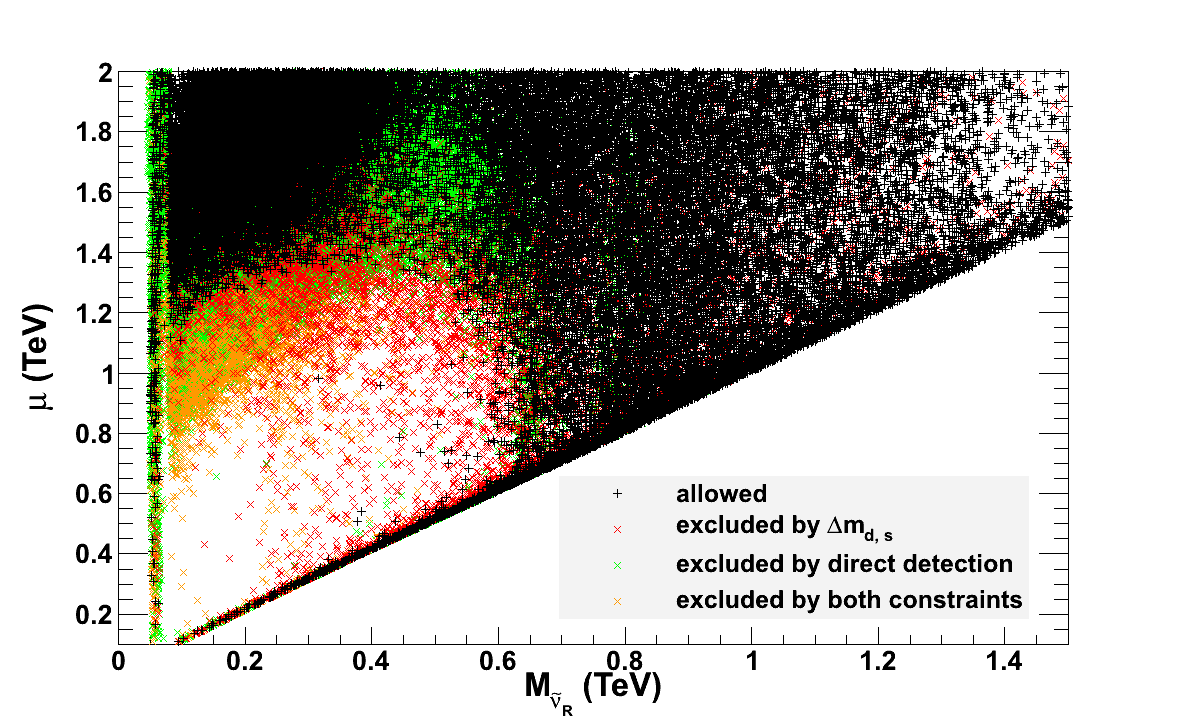}
\ablabels{63}{144}{30} 
  \caption{The allowed $U(1)_\eta$ scenarios in the   a) $M_{Z_2}$ vs $m_{\lsp}$ plane and 
    b)  $\mu$ vs $m_{\lsp}$ plane. In red,  points excluded by the $\Delta M_{d,s}$ constraints, in green those excluded by XENON100 
    and in yellow those excluded by both constraints.}\label{fig:random_eta}
\end{figure}

We also explore the parameter space of the model, varying the parameters in the range shown in Table~\ref{tab:range}.
As before, we impose in addition to limits on $\Omega h^2$ and on the SI cross section,  the lower limit on the Higgs and $Z_2$ mass, and the limit on  $\Delta M_{d,s}$.
The results are displayed in Fig.~\ref{fig:random_eta} in the $M_{Z_2}-\mlsp$ plane, as well as in the $\mu-\mlsp$ plane. 
As we have discussed above, because Higgs annihilation is efficient enough, the sneutrinos  are not confined to the $h_i,Z_2$ resonance region.
Sneutrinos from 100GeV to $M_{Z_2}/2$ can satisfy the relic density constraint, either through pair annihilation or through 
coannihilation (the region where $\mlsp\approx \mu$). In the former case large values of $\mu$ are preferred to have large enough couplings of the sneutrino LSP to $h_1$.   
The direct detection cross section is large with a maximum value near $\ssi \approx 10^{-7}$~pb for the whole range of LSP masses. 
Because the experimental limits on the direct detection rate are more stringent for masses around 80 GeV, light sneutrinos are severely constrained especially when the $Z_2$ mass is lighter than 2TeV. 
The lower bound on the cross section   for $M_{Z_2}\leq 3$~TeV is  $\ssi=2.\times 10^{-9}$~pb . 
The typically large direct detection rate is the main signature of this scenario. To completely probe this scenario 
up to DM masses of 0.3(1.5) TeV requires one (two) orders of magnitude improvement  over the current limits. 
Several of the scenarios are also constrained by $\Delta M_s$, in particular when $\alpha_Z<0$ which implies $\tan\beta
<1$. This does not affect the range of values of sneutrino masses compatible with all constraints.

\begin{figure}[!htb]
\includegraphics[width=8cm,height=6.5cm]{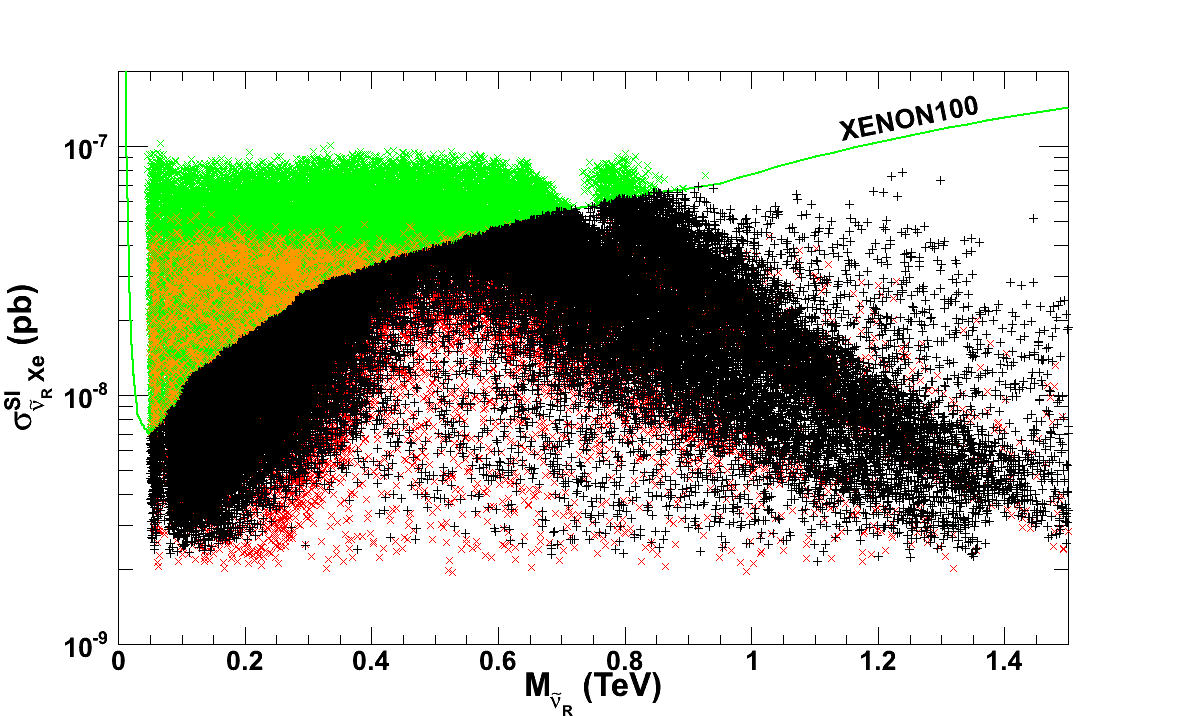} 
\includegraphics[width=8cm,height=6.5cm]{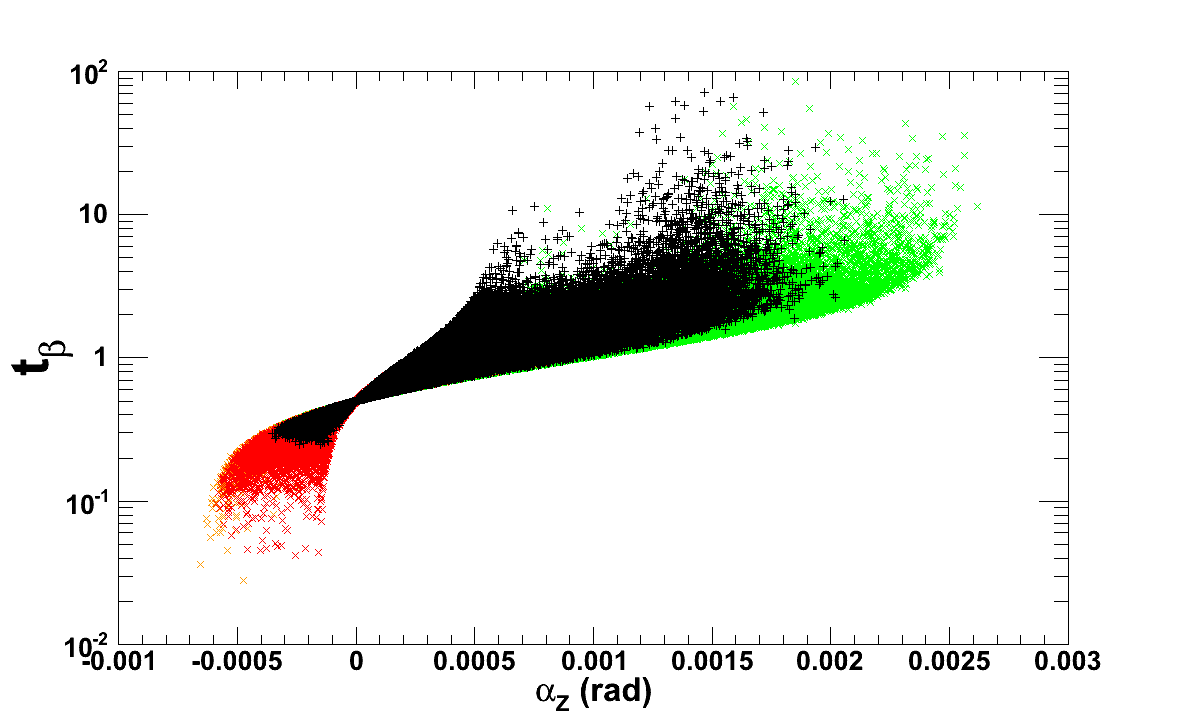}
\ablabels{63}{144}{15} 
  \caption{a)
$\sigma^{SI}_{\lsp Xe}$ as a function of the LSP mass for the allowed scenarios in the $U(1)_\eta$ model.
b) Allowed scenarios in the $\tan\beta$- $\alpha_Z$ plane, same color code as Fig.~\ref{fig:random_eta}.  }
\label{fig:random_eta_direct}
\end{figure}

\subsection{A global scan of the parameter space}

Having illustrated the properties of the sneutrino dark matter in specific models we will next
explore the complete parameter space of the model by keeping all parameters for the neutralino, sfermion and gauge boson sector free
while imposing the constraints from WMAP, direct detection as well as on Higgs masses. 

\begin{figure}[!ht]
  \centering
\includegraphics[width=10cm]{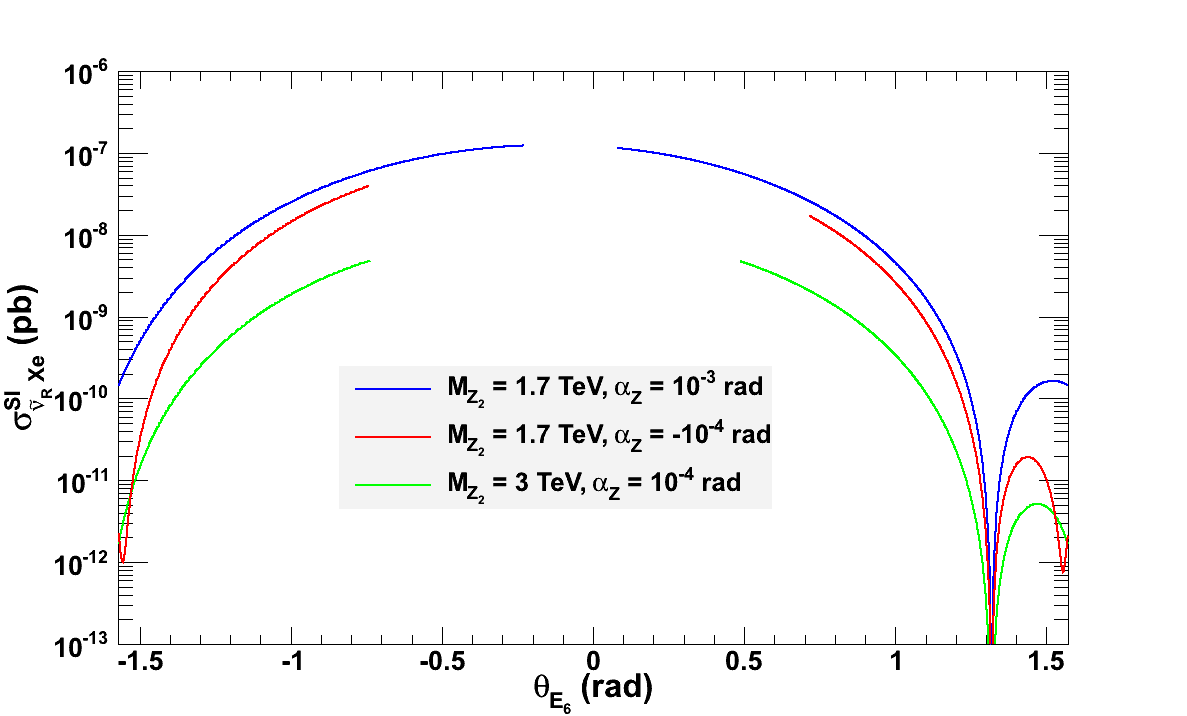}
   \caption{$\ssi$ as a function of $\tesix$ for
  $(\mzp,\azz)=
   (1.7,10^{-3}),(1.7,-10^{-4}),(3,10^{-4})$, $\mu=M_{Z_2}/2+0.5$~TeV, $A_\lambda=1$~TeV and $\mlsp=M_{Z_2}/2$.  }
\label{fig:sigmae6}
\end{figure}

The values of $\theta_{E_6}$ where the sneutrino is a good dark matter candidate are restricted. 
First for  $\tesix \approx 0$ the value of  $v_s$ becomes very large especially when $M_{Z_2}$ is large. 
This induces large negative corrections to sfermion masses and lead in particular to a charged LSP.
For example, for soft terms at 2TeV the values $-0.2<\theta_{E_6}<0.05$ are excluded when $M_{Z_2}=1.$TeV.
Second the direct detection cross section is often too large when $|\tesix|<0.5$. 
This is due mainly to the contribution of the $Z_2$ exchange to the direct detection cross section that is proportional to $\cos\tesix$.
To illustrate this we display the variation of the direct detection cross section as a function of $\tesix$ for different values of $M_{Z_2}$ in Fig.~\ref{fig:sigmae6}.
We choose $\mlsp=M_{Z_2}/2$ so as to guarantee that the WMAP upper bound is satisfied and fix $\mu=M_{Z_2}/2+0.5$~TeV in each case to ensure that the sneutrino is the LSP. We have also fixed $M_{\tilde f}=2$~TeV. The direct
detection bound are easily satisfied near $\tesix=\pm \pi/2$ because of the suppressed contribution of the Z vectorial coupling, see
Fig.~\ref{fig:sigmae6}.
The mixing angle $\azz$ has only a moderate impact on the direct detection rate, while increasing the mass of the $M_{Z_2}$ reduces the SI cross section, 
except when $\tesix=\theta_\psi$ as we have seen in Section~\ref{sec:psi}. Note that for $\alpha_z<0$ there is a dip in the cross section near $\tesix=\pi/2$, this is because of a cancellation between the $y$ and $y'$ contributions in Eq.~\ref{eq:ddz} in the cross section on neutrons.

\begin{figure}[!ht]
\includegraphics[width=8cm,height=6.5cm]{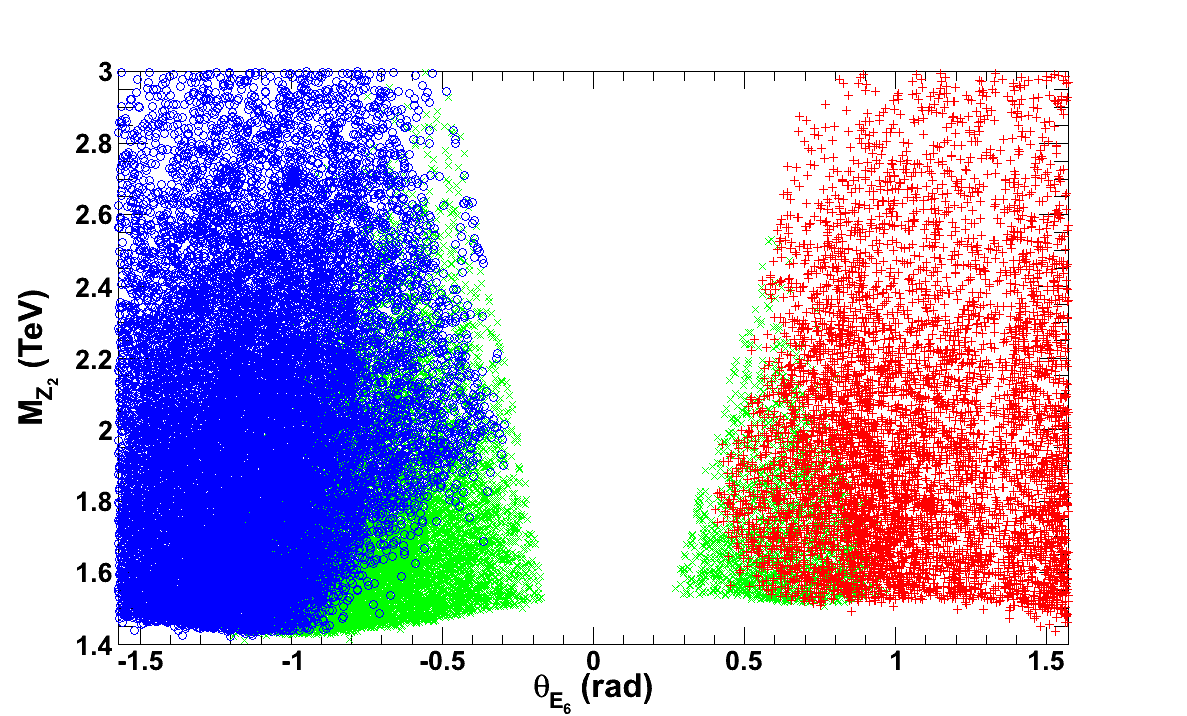}  
\includegraphics[width=8cm,height=6.5cm]{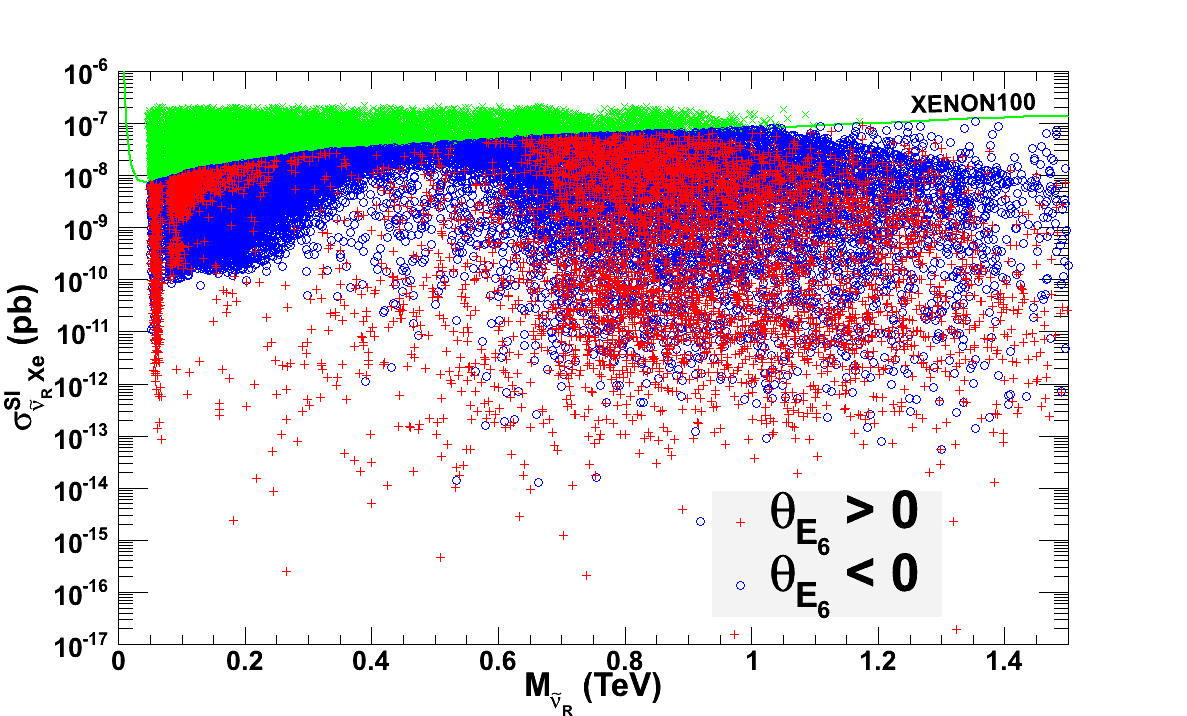} 
\ablabels{38}{92}{15}
  \caption{  a) Allowed scenarios in the $M_{Z_2}$ vs $\tesix$ plane, the models above the XENON100 bound are in green (light grey). 
  b) $\ssi$ as a function of $\mlsp$.   }
\label{fig:te6:sigma}
\end{figure}

For the random scans we choose the same range as in Table~\ref{tab:e6} with in addition $-\pi/2 \leq \tesix \leq \pi/2$.
As before, we have applied the constraints on $M_{Z_2}$ from the LHC and on $\alpha_Z$. 
 We have imposed the latest constraint from XENON100~\cite{Aprile:2010um} and from $\Delta M_s$ a posteriori to 
better illustrate the impact of these constraints.
The  successful scenarios   are displayed in Fig.~\ref{fig:te6:sigma} in the $M_{Z_2}-\tesix$ plane. 
The white region at small values of $\tesix$ have a charged fermion LSP.  Many scenarios with 
$-\pi/4<\tesix<0 $ have too large a direct detection rate as was the case for the model $U(1)_\eta$. The value of the direct detection cross section 
span several orders of magnitude from less than $10^{-13}$~pb to $2\times 10^{-7}$~pb. In particular for sneutrino masses around 100 GeV there are many models 
which exceed the direct detection limit. These are scenarios with $\tesix<0$ where the sneutrino coupling to light Higgs allows
efficient annihilation even away from resonance as discussed in the case of $U(1)_\eta$. The enhanced couplings to the Higgs  as well as the coupling 
to the $Z_{1,2}$ lead to a large direct detection rate. Models with $\tesix>0$ have only light sneutrinos in that mass range when
 coannihilation plays a role and therefore tend to have much lower direct detection rate. 
In fact even for heavier masses models with $\tesix>0$ predict
 smaller direct detection rates as discussed above.

\begin{figure}[!ht]
\includegraphics[width=8cm,height=6.5cm]{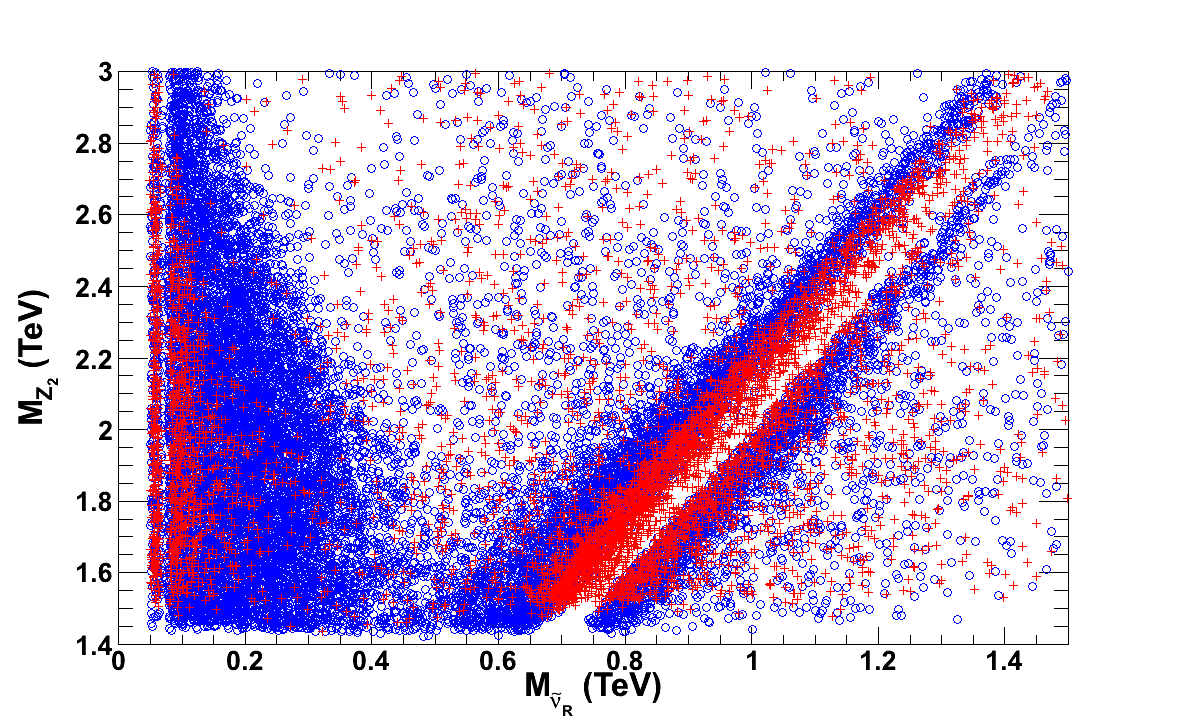} 
\includegraphics[width=8cm,height=6.5cm]{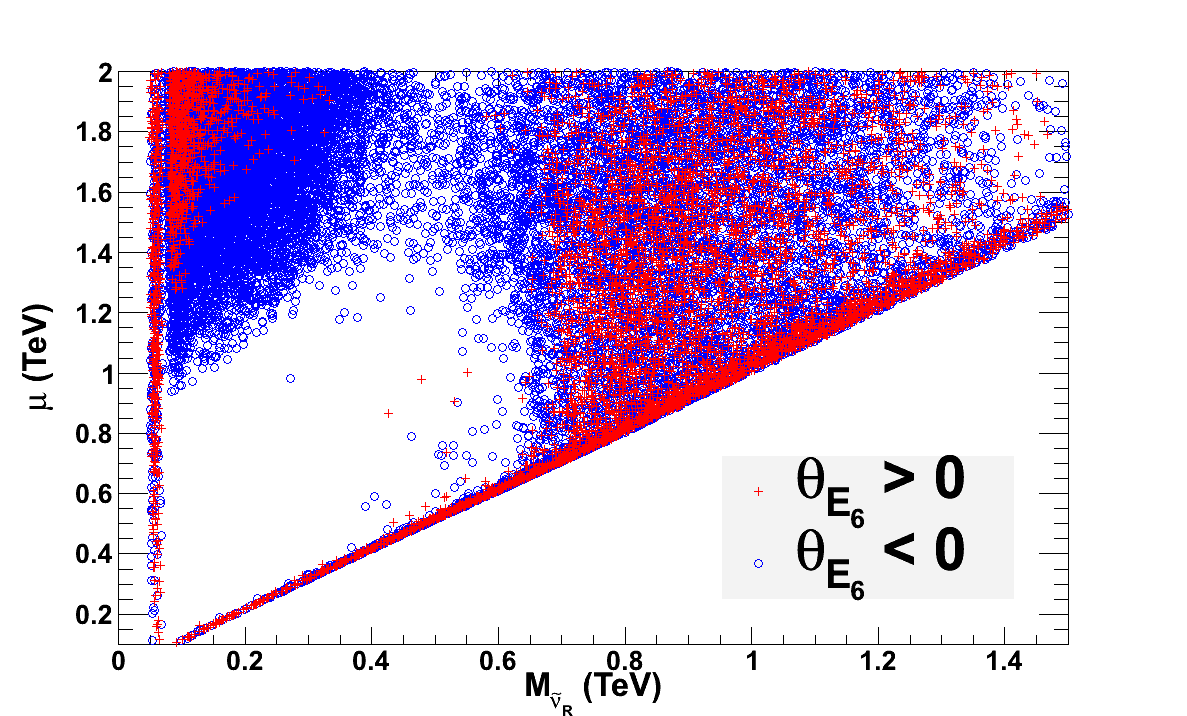}
\ablabels{63}{144}{15}
  \caption{  The allowed scenarios in the   a) $M_{Z_2}$ vs $m_{\lsp}$ plane and 
    b)  $\mu$ vs $m_{\lsp}$ plane after applying all constraints.}
\label{fig:e6}
\end{figure}

The processes that can contribute to $\Omega h^2$ were discussed both in the context of $U(1)_\psi$ and $U(1)_\eta$ models.
In the general case we find similar results. We find a predominance of annihilation near a $h_1$ or  singlet Higgs/$Z_2$ resonance as well as annihilation into gauge boson pairs through $h_1$ exchange. The latter being confined to sneutrino masses just above the W pair threshold when $\tesix>0$. These regions have a high density of points  in the plane $M_{Z_2}-\mlsp$ in  Fig.~\ref{fig:e6}a.  For $\tesix>0$ the only other allowed scenarios have $\mlsp \approx
 \mu$ as displayed in Fig.~\ref{fig:e6}b. These are dominated by Higgsino coannihilation.  
  For $\tesix<0$, sneutrinos of masses above 100 GeV can also annihilate efficiently through $h_1$ exchange provided their coupling to $h_1$ is large enough - this requires large values of $\mu$. In both figures we have imposed the $\Delta M_{d,s}$ constraint although we do not display explicitly its impact. As discussed in the previous section this constraint plays a role for small values of $\tan\beta$ for any values of the LSP mass.
  
\section{Conclusion}

The right-handed sneutrino is a  viable thermal dark matter candidates in U(1) extensions of the MSSM. 
The allowed parameter space depends strongly on the value of the new $Z_2$ vector boson mass.  Sneutrino annihilation is typically dominated by resonance annihilation,   with in 
particular the dominant contribution from the  Higgs sector  rather than  from the new gauge boson $Z_2$.
 For the light Higgs this requires fine tuning of the masses while for the heavy singlet Higgs the
mass difference has to be within 15\% (of $m_{h_i}/2$). As in other supersymmetric models, coannihilation processes can also be important. 
For simplicity we have only discussed the case of one sneutrino dark matter candidate, in the case of complete degeneracy of three sneutrinos the relic
density increases by a factor 3. Indeed the annihilation cross section is the same for all sneutrino flavours since there is never a significant
contribution from annihilation into neutrino pairs, thus the increase in the number of channels is compensated by the increase in the degrees of
freedom leading to a smaller effective annihilation cross section. This will imply a narrower range of masses  for the LSP near $M_{h_i}/2$. 

The direct detection limit is very stringent for a whole class of models  unless the mass of the $Z_2$ is above 2TeV.
These scenarios  with $\tesix<0$ will be best tested with SI detectors with improved sensitivities. In particular with a factor of 2 better sensitivity the whole region with intermediate masses for the sneutrinos - when  annihilation into W pairs is dominant - will be probed. The scenarios with $\tesix>0$ are more challenging to probe via direct detection. Another  signature of the RHSN model is that of a new gauge boson at the LHC. In fact the recent results from searches for $Z_2$ gauge bosons at the LHC have had  a significant  impact on the parameter space of the model and pushed one of the favored region for the sneutrino LSP to be near 800GeV. 
A $Z_2$ up to several TeV's will be probed when the energy of the LHC is increased to 14TeV. 
A negative result from such a search will imply a much reduced rate for direct detection searches as well. 
Indirect searches could also provide a good probe of the RHSN model, these will be investigated separately. 

In this analysis we have assumed that Dirac neutrinos were very light (sub-eV range).
Such right-handed Dirac neutrinos can be produced by Z' interactions prior to big bang nucleosynthesis leading to a faster expansion rate of the universe and to too much $He^4$. The resulting constraints on the mass of the new gauge bosons in the $E_6$ model were analyzed in ~\cite{Barger:2003zh} and compatibility with BBN resulted into lower limits on the Z' mass in the multi-TeV range assuming the effective number of neutrinos was increased by 0.3. 
However since then new data from WMAP have been released which implied a higher than expected value  for the abundance of $He^4$
and for the effective number of relativistic neutrino species, more precisely $N_{eff}=4.34+0.86-0.88$~\cite{Komatsu:2008hk,Krauss:2010xg}. This new value would have the effect of relaxing the lower limit on the Z' mass to a value comparable to LHC bounds.

\section{Acknowledgements}
G.B. thanks the LPSC, Grenoble for its hospitality. We thank S. Viel for discussions on the LHC bounds on new gauge bosons.
This work was supported  by the GDRI-ACPP of CNRS.
The work of AP was supported by the Russian foundation for Basic Research, 
grant RFBR-10-02-01443-a. 

\appendix\label{sec:app}
\section{The masses in the Higgs sector}

The tree-level CP-even Higgs mass matrix in 
 the basis
($H_d^0, H_u^0, S$) reads~\cite{Barger:2006dh}
\begin{align}
\left({\mathcal{M}_{+}^0}\right)_{11} & =  \left[\frac{({g_Y}^2 + g^2_2)^2}{4} +  Q^2_1 {g'_1}^{2}\right] (v c_\beta)^2 + \frac{\lambda A_\lambda t_\beta v_s}{\sqrt{2}}\nonumber \\
\left({\mathcal{M}_{+}^0}\right)_{12} & = -\left[\frac{({g_Y}^2 + g^2_2)^2}{4} - \lambda^{2} - Q_1 Q_2 {g'_1}^{2}\right] v^2 s_\beta c_\beta - \frac{\lambda A_\lambda v_s}{\sqrt{2}}\nonumber\\
\left({\mathcal{M}_{+}^0}\right)_{13} & =  \left[\lambda^{2} + Q_1 Q'_S {g_{1'}}^{2}\right] v c_\beta v_s - \frac{\lambda A_\lambda v s_\beta}{\sqrt{2}}\nonumber\\
\left({\mathcal{M}_{+}^0}\right)_{22} & =  \left[\frac{({g_Y}^2 + g^2_2)^2}{4} + Q^2_2 {g'_1}^{2}\right] (v s_\beta)^2 + \frac{\lambda A_\lambda v_s}{t_\beta \sqrt{2}}\nonumber \\
\left({\mathcal{M}_{+}^0}\right)_{23} & =  \left[\lambda^{2} + Q_2 Q'_{S} g^2_1\right] v s_\beta v_s - \frac{\lambda A_\lambda v c_\beta}{\sqrt{2}}\nonumber \\
\left({\mathcal{M}_{+}^0}\right)_{33} & =  Q'^2_S g'^2_1 v^2_s + \frac{\lambda A_\lambda v^2 s_\beta c_\beta}{v_s \sqrt{2}}, 
\end{align}

The dominant radiative corrections due to top and stop are 
\begin{align}
({\mathcal{M}_{+}^1})_{11} & = k \left[ \left( \frac{({\widetilde m}^2_1)^2}{(m_{\widetilde t_1}^2 - m_{\widetilde t_2}^2)^2} {\mathcal G} \right) (v c_\beta)^2 + \left( \frac{\lambda h_t^2 A_t}{2 \sqrt{2}} \mathcal{F} \right) t_\beta v_s \right]\nonumber \\
({\mathcal{M}_{+}^1})_{12} & = k \left[ \left( \frac{{\widetilde m}^2_1 {\widetilde m}^2_2}{(m_{\widetilde t_1}^2 - m_{\widetilde t_2}^2)^2} {\mathcal G} + \frac{h_t^2 {\widetilde m}^2_1}{m^2_{\widetilde t_1} + m^2_{\widetilde t_2}} (2-{\mathcal G}) \right) v^2 s_\beta c_\beta -  \left( \frac{\lambda h_t^2 A_t}{2 \sqrt{2}} \mathcal{F} \right)s \right]\nonumber \\
({\mathcal{M}_{+}^1})_{13} & = k \left[ \left(\frac{{\widetilde m}^2_1 {\widetilde m}^2_s}{(m_{\widetilde t_1}^2 - m_{\widetilde t_2}^2)^2} {\mathcal G} + \frac{\lambda^2 h_t^2}{2} {\mathcal F} \right) v c_\beta v_s - \left( \frac{\lambda h_t^2 A_t}{2 \sqrt{2}} \mathcal{F} \right) v s_\beta \right]\nonumber \\
({\mathcal{M}_{+}^1})_{22} & = k \left( \frac{({\widetilde m}^2_2)^2}{(m_{\widetilde t_1}^2 - m_{\widetilde t_2}^2)^2} {\mathcal G} + \frac{2 h_t^2 {\widetilde m}^2_2}{m^2_{\widetilde t_1} + m^2_{\widetilde t_2}} (2-{\mathcal G}) + h_t^4 \ln \frac{m^2_{\widetilde t_1} m^2_{\widetilde t_2}}{m_t^4} \right) (v s_\beta)^2 \nonumber\\
& \quad + k \left( \frac{\lambda h_t^2 A_t}{2 \sqrt{2}} \mathcal{F} \right) \frac{v_s}{t_\beta}\nonumber \\
({\mathcal{M}_{+}^1})_{23} & = k \left[ \left(\frac{{\widetilde m}^2_2 {\widetilde m}^2_s}{(m_{\widetilde t_1}^2 - m_{\widetilde t_2}^2)^2} {\mathcal G} + \frac{h_t^2 {\widetilde m}^2_s}{m^2_{\widetilde t_1} + m^2_{\widetilde t_2}} (2-{\mathcal G}) \right) v s_\beta v_s - \left( \frac{\lambda h_t^2 A_t}{2 \sqrt{2}} \mathcal{F} \right)v c_\beta \right]\nonumber\end{align}\begin{align}
({\mathcal{M}_{+}^1})_{33} & = k \left[ \left(\frac{({\widetilde m}^2_s)^2}{(m_{\widetilde t_1}^2 - m_{\widetilde t_2}^2)^2} {\mathcal G} \right) v_s^2 +  \left( \frac{\lambda h_t^2 A_t}{2 \sqrt{2}} \mathcal{F} \right) \frac{v^2 s_\beta c_\beta}{v_s}\right], \label{eq8}\end{align}

where $k ={3/(4\pi)^2}$, $h_t$ is the top Yukawa coupling and
\begin{align}
{\cal G} & = 2\left[1- \frac{m^2_{\tilde t_1}+ m^2_{\tilde t_2}}{m^2_{\tilde t_1}- m^2_{\tilde t_2}} \log \left( {m_{\tilde t_1} \over m_{\tilde t_2}} \right)\right] 
\nonumber\\
{\cal F} & = \log \left( \frac{m^2_{\tilde t_1} m^2_{\tilde t_2}}{\Lambda^4}\right) - {\cal G}(m^2_{\tilde t_1}, m^2_{\tilde t_2})
\end{align}
and
\begin{align}
\widetilde{m}_1^{2} & = h_t^2 \mu \left(\mu - A_t t_\beta \right)\nonumber\\
\widetilde{m}_2^{2} & = h_t^2 A_t \left(A_t - \frac{\mu}{t_\beta} \right)\nonumber\\
\widetilde{m}_s^{2} & = \left(\frac{v c_\beta h_t}{v_s} \right)^2 \mu (\mu - A_t t_\beta)\nonumber\\
\end{align}
The $\overline{{\rm DR}}$ renormalization scale, $\Lambda$ is fixed to 1 TeV. 
The CP-odd mass tree-level matrix reads
\begin{align}
\left({\mathcal{M}_{-}^0}\right) & =\frac{\lambda A_\lambda}{\sqrt{2}} \begin{pmatrix}
    t_\beta v_s & v_s & v s_\beta\\
    v_s & \frac{v_s}{t_\beta} & v c_\beta\\
    v s_\beta & v c_\beta & \frac{v^2 s_\beta c_\beta}{v_s}
  \end{pmatrix} \\
\end{align}
and the one-loop corrections due to top and stop
\begin{equation}
(\mathcal{M}^1_{-})_{ij}  = \frac{ \lambda v^2 s_\beta c_\beta v_s}{\sqrt 2 v_i v_j}\frac{k h_t^2 A_t}{2} {\cal F}, \quad i,j \in \{1,2,3\}, \label{eq9}
\end{equation}
where $v_1=v_d,v_2=v_u,v_3=v_s$.

The charged Higgs mass including radiative corrections from the top and stop sector,
\begin{eqnarray}
m^2_{H^\pm} & =& \frac{\lambda A_\lambda \sqrt{2}}{\sin 2 \beta} v_s - \frac{\lambda ^2}{2} v^2 +  M^2_W + \Delta^2_{H+} \nonumber\\
\Delta^2_{H+}& =& {\lambda A_t v_s k h_t^2 {\cal F}\over \sqrt 2 \sin 2 \beta} + {3 {g_2}^2 \over 32 \pi^2 M_W^2}\left({2 m_t^2 m_b^2\over s^2_\beta c^2_\beta}-M_W^2\left({m_t^2\over s^2_\beta}+{m_b^2\over c^2_\beta}\right)+{2\over3}M_W^4\right) \log{ \Lambda^2\over m_t^2}. \label{eq10}
\end{eqnarray}
where the SUSY scale $\Lambda$ is taken to be 1 TeV.

\bibliography{umssm}{}

\end{document}